\documentclass[final]{IEEEtran}
\pdfoutput=1
\usepackage{graphicx}
\usepackage{subfigure}
\usepackage[cmex10]{amsmath}
\usepackage{amssymb}
\newcommand{\be}{\vspace{-0.03cm}\begin{eqnarray}}
\newcommand{\ee}{\end{eqnarray}\vspace{-0.03cm}}

\newcommand{\stbl}[1]{\setlength{\tabcolsep}{0.23em}#1}
\usepackage{array}
\newcolumntype{M}[1]{@{}>{\centering\arraybackslash}m{#1}@{}}
\newcolumntype{N}[1]{>{\raggedright\arraybackslash}m{#1}}
\newcolumntype{P}[1]{>{\raggedleft\arraybackslash}m{#1}}
\newcolumntype{C}[1]{@{}>{\centering\arraybackslash}p{#1}@{}}
\newcolumntype{L}[1]{>{\raggedright\arraybackslash}p{#1}}
\newcolumntype{R}[1]{>{\raggedleft\arraybackslash}p{#1}}
\usepackage{algorithmic}

\usepackage{xcolor}

\title{Performance Improvement of Time-Balance Radar Schedulers Through Decision Policies (Extended Version)}
\author{\"Omer \c{C}ay{\i}r, \c{C}a\u{g}atay Candan
\\ Department of Electrical and Electronics Engineering, METU, Ankara, Turkey
\\ \{ocayir, ccandan\}@metu.edu.tr}

\def\Nk{N_k}
\def\xrq{x_{\text{requesting},k}}
\def\nonu{\IEEEnonumber\\}
\def\sbjct{\operatorname{subject\phantom{.}to}}
\def\sbjctb{\phantom{m}\!}
\def\sbjcts{\sbjctb\phantom{\sbjct}}

\def\vnupd{V^n_{nupd}}
\def\vupd{V^n_{upd}}
\def\vf{V^n_{}}
\def\thtz{\theta_0}
\def\thto{\theta_1}

\def\infst{\mu^n_k}
\def\infstn{\mu^n_{k+1}}
\def\clH{{\cal H}}
\def\cHb{\clH_{bt}}
\def\cHg{\clH_{gt}}
\def\Hbt{\cHb^{}}
\def\Hgt{\clH_{gt}^{}}
\def\clP{{\cal P}}
\def\Pbt{\clP_{bt}^{}}
\def\Pgt{\clP_{gt}^{}}
\def\thM{\mu_{th}^n}

\def\Knk{{\cal K}^n_k}
\def\anM{{\cal A}_M(r)}
\def\bnM{{\cal B}_M(r)}

\def\tmb{t_{\text{TB}}}
\def\kxx{\times{10}}
\def\kx{\!\!\times\!\!{10}}

\def\uknz{u^n_k=\text{NUPD}}
\def\ukno{u^n_k=\text{UPD}}
\def\nmv{us}
\def\mv{ss}

\newcommand{\hh}{\hline\hline}
\newcommand{\btbdc}{\begin{table*}[!t]\centering\renewcommand{\arraystretch}{1.3}}
\newcommand{\etbdc}{\end{table*}}
\newcommand{\btb}{\begin{table}[!t]\centering\renewcommand{\arraystretch}{1.3}}
\newcommand{\etb}{\end{table}}
\newcommand{\bmip}{\hh \begin{minipage}{\columnwidth} \vspace{.5em}}
\newcommand{\emip}{\vspace{.5em} \end{minipage}\\ \hh}
\newcommand{\bea}{\setlength{\arraycolsep}{0.0em}\begin{eqnarray}}
\newcommand{\eea}{\end{eqnarray}\setlength{\arraycolsep}{5pt}}
\newcommand{\eql}{&{}={}&}
\newcommand{\teql}{&{}\triangleq{}&}
\DeclareMathOperator{\trace}{tr}
\usepackage[pdftex,pdfusetitle]{hyperref}
\hypersetup{
	bookmarksopen,
	bookmarksdepth= 2,
	pdfinfo={
		Subject={The resource management of a phase array system capable of multiple target tracking and surveillance is critical for the realization of its full potential. Present work aims to improve the performance of an existing method, time-balance scheduling, by establishing an analogy with a well-known stochastic control problem, the machine replacement problem.},
		Keywords={Multi-Function Radar, Radar Resource Management, Radar Task Scheduling, Time-Balance Schedulers, Machine Replacement Problem.}
	}
}
\begin{document}
\maketitle
\begin{abstract}
The resource management of a phase array system capable of multiple target tracking and surveillance is critical for the realization of its full potential. Present work aims to improve the performance of an existing method, time-balance scheduling, by establishing an analogy with a well-known stochastic control problem, the machine replacement problem. With the suggested policy, the scheduler can adapt to the operational scenario without a significant sacrifice from the practicality of the time-balance schedulers. More specifically, the numerical experiments indicate that the schedulers directed with the suggested policy can successfully trade the unnecessary track updates, say of non-maneuvering targets, with the updates of targets with deteriorating tracks, say of rapidly maneuvering targets, yielding an overall improvement in the tracking performance.
\end{abstract}



\setcounter{page}{1}
\section{Introduction}
A modern radar system is required to handle a variety of tasks, such as surveillance, multi-target tracking, calibration, guidance etc. The capabilities of such a system, say a multi-function radar system, come at a significant initial deployment cost mainly due to the installment of possibly thousands of transmit/receive modules. Taking the full advantage of the mentioned capabilities requires an effective radar resource management (RRM). Typically, the multitude of tasks in execution compete for the radar resources, namely time, energy and computation \cite{ding08,dingbook,Keuk1993}. In this work, we focus on the time allocation problem for such systems.

The allocation of time, among other resources, is generally called \textit{scheduling} in the RRM applications. Scheduling methods can be classified into two classes, adaptive and non-adaptive methods \cite{ding08}. Non-adaptive scheduling methods, namely heuristic schedulers, are based on a rule-based design. The behavior of schedulers and prioritization (priority assignment) of tasks are pre-defined by the fixed rules. In contrast, the adaptive scheduling methods dynamically determine the task prioritization and scheduling to optimize overall performance. According to the importance of tasks being scheduled, an efficient task prioritization process is required to rank tasks for the performance improvement of both adaptive and non-adaptive schedulers. Knowledge-based systems using some a-priori information is suggested to this aim.  Knowledge-based systems consist of two sub-systems, a knowledge database containing information related to the system environment and an inference engine making final decisions taking into account both a-priori information and existing conditions \cite{ginibook}. In \cite{miranda}, a fuzzy logic based approach is suggested to rank targets and surveillance sectors for dynamically changing system environments. For tracking tasks, the priorities are assigned according to five different fuzzy variables such as quality of tracking, hostility, degree of threat. For surveillance tasks, there are four fuzzy variables including the original priority and number of threatening targets. In \cite{komorniczak,komorniczak2}, a neural network based approach is utilized for target ranking with respect to range, radial velocity, membership (friend or foe), acceleration and object rank (important or not important). Both neural network and fuzzy logic based approaches provide an adaptive priority assignment, in general. Especially, the learning capability of neural network based schedulers enable the operator to update the system behavior after the detection of new targets. However, the learning process is far from trivial \cite{dingbook}. The process includes training of several data sets in random order from the same initial starting point \cite{komorniczak}.

In this work, we aim to achieve the benefits of adaptive scheduling without a major sacrifice from the low computational load of non-adaptive scheduling methods. More explicitly, the adaptive scheduling schemes in the literature are based on the stochastic control and they are, in general, difficult to implement due to high computational requirements, \cite{vikram1, vikram2,vikram2c}. To reap the benefits of adaptive scheduling while maintaining a low computational load, we suggest an improvement over a well-known non-adaptive scheduling scheme, namely the time-balance scheduler, based on a classical stochastic control problem, namely the machine-replacement problem. The suggested improvement, in effect, yields to an automated task prioritization and shown to have good adaptation capabilities to target tracking scenario unfolding to the operator.

The time-balance (TB) method is based on the idea of meeting the ``deadlines'' of each task with the minimum possible delay. The TB method is robust and achieves the desired task occupancies in the long horizon. In Section~\ref{sec:tb}, the general features of TB schedulers are further described. A similar adaptation effort on the TB method is the adaptive task prioritization, as described in \cite{torres2}, with the aim of completing the surveillance task properly even when radar is overloaded with the task of tracking a large number of targets and without adjusting the task update times. With this method, the task prioritization adapts to a predictable task queue and is independent of radar performance measurements, the track scores.

In this work, we aim to develop a target selection procedure for the TB method in order to reduce the tracking error, when the radar system is overloaded with tracking tasks having identical priorities. The proposed method is based on a well-known stochastic control problem known as the machine replacement problem, as given in \cite{grosfeld13}. Here, our goal is to construct an analogy between the well-known control problem and the target tracking problem and enable the utilization of the results for this problem in the performance improvement of the TB schedulers. The suggested method and its variants are highly practical and can be immediately applied in the existing systems utilizing the conventional TB schedulers.

\section{Background: Time-Balance Method and Machine Replacement Problem}\label{sec:tb}

\subsection{Time-Balance Method}
The time-balance metric gives the degree of urgency of each task during the radar operation. A revisit time is assigned to each task and the TB metric is continually updated to reflect the approaching visitation deadline of each task, \cite{stafford,wray}. More specifically, each task is associated with a TB value, $\tmb$. A positive $\tmb$ value indicates an overdue task. A negative $\tmb$ value indicates a task whose immediate execution would be ahead of the assigned deadline. A zero $\tmb$ value indicates a just-on-time task. At any time, a new task can be inserted to the list by assigning a negative $\tmb$ value. If a task is scheduled, its $\tmb$ is decreased by its task update time (revisit time). Upon execution of any task, the $\tmb$ of other tasks which are not scheduled is increased by a fixed amount determined by the designer. Under light load conditions, the TB method is highly efficient enabling timely task updates. As the load increases, the TB method suffers a performance loss since this method does not have any capacity to discriminate tasks as urgent and not-so-urgent.

A scheduler algorithm that utilizes the TB method is employed in the Multifunction Electronically Scanned Adaptive Radar (MESAR), \cite{stafford,butlerphd}. This method allows to divide tasks into subtasks (looks) that can be interleaved to manage radar time efficiently and decrease the delays for the highly prioritized tasks by starting from the highest priority level at each scheduling instant. In \cite{torres}, the TB scheduler chooses the task which has higher $\tmb$ than other tasks as the next task. The scheduler is designed to schedule mainly the tracking tasks, and the surveillance task is fragmented by the task fragment time. That is, the surveillance task is not periodically started, but one of its fragments is scheduled whenever all tracking tasks have negative $\tmb$ value, i.e. when there is some idle time between the tracking tasks.

The adaptive time-balance (ATB) scheduler is proposed in \cite{torres2}. Here, the surveillance task is associated with a $\tmb$ value so that it is scheduled with respect to task update time to detect new targets. The task update times can be adaptively changed as a possible solution for the overload conditions or to increase the revisit improvement factor. The ATB scheduler supports user defined priority levels for each task, and tasks are scheduled according to these priority levels and $\tmb$'s.

In this work, we present a further improvement on the TB method. Our goal is to schedule the target tracking tasks according to the track quality. Hence, we would like to have adjust the scheduling parameters of the TB method dynamically according to the unfolding tracking scenario.

The target selection problem emerges when there are more than one target requesting the track update. The conventional TB scheduler follows the steps below:
\begin{enumerate}
\item Select the targets with the highest priority level,
\item Look for the targets which have the highest $\tmb$.
\end{enumerate}

Typically, if there are multiple overdue tracking tasks at the same priority level, the executed task is selected according to the first-come, first-serve (FCFS) principle, \cite[Chapter 6]{silberschatzbook}. This method aims to minimize the overall lateness in the task execution. Clearly, this is an efficient mechanism for the maneuvering targets requiring a rapid execution depending on the type of maneuver. Our goal is to include the information about track quality in the task selection. To this aim, we construct an analogy between the well-known machine replacement problem and RRM problem.

\subsection{Machine Replacement Problem}
We describe the machine replacement problem with a concrete example. Assume a baker having the main asset of an oven (machine) which can be either in ``good'' or ``bad'' state related to its cooking performance. The state of the machine deteriorates due to aging, and the products of the machine can be delicious (conforming) or tasteless (defective) depending on the state variable. The true state of the oven is not known, but can be observed by the quality of the products. It is possible to have a bad product in spite of the good state of the machine with a non-zero probability and vice versa. In this problem, it is assumed that the cost of a new machine (replacement cost), price for the good and bad products are fixed quantities. The main question is to determine the time to replace the machine yielding the profit maximization. This type of optimization problems is categorized as partially observable Markov decision processes (POMDPs). Here, the Markov process is related with unknown state of the machine that can be only observed in the presence of noise \cite{vikrambook}.

We establish an analogy between the resource management problem and machine replacement problem as follows: For the target selection problem, a target can be in one of two states, namely up-to-date and stale. The up-to-date state denotes that the target track is predictable with a high accuracy by the tracker and may not require an immediate track update. Hence, the up-to-date state can be considered the ``good'' state. The stale state denotes that the target track is not predictable with a good accuracy and this track may require the urgent attention of the scheduler due to its higher probability of target drop. Hence, it is the ``bad'' state. As in the POMDP problems, we have noisy information on the target states.

As discussed in the latter parts of this paper, we assign a state to each target and utilize the track quality information as the noisy measurement on the state. The proposed analogy is especially valuable for an overloaded scheduler; but, even for the underloaded case, it can yield some performance improvements. It should be noted that an unnecessary execution of a target update in the up-to-date state could decrease the tracking performance of other targets. Hence, the state dependent track update selection can also be beneficial in the improvement of overall track quality.

\section{Proposed Machine Replacement Problem Based Policy}
We utilize several results, with some corrections, from the work of T. Ben-Zvi, and A. Grosfeld-Nir, \cite{grosfeld13}. Here, the binomial observation model for the machine replacement problem refers to the classification of the quality of products as \textit{conforming units} or \textit{defective units} according to observations (measurements) while the production process is in either ``good'' or ``bad'' state. The true state of the process is not observable and can only be estimated with some error. Thus, the production process is modeled as a POMDP with some control limits. The POMDPs are known to be usually hard to solve due to prohibitively large size of the state space \cite{grosfeld06}. In \cite{grosfeld13}, it is proven that the infinite-horizon control limit defined as a function of the probability of obtaining a conforming unit can be calculated by solving a finite set of linear equations.

In the target selection problem, there are many targets and each target is, conceptually, associated with a machine. At each instant of decision-making, a target is selected among a set of overdue targets according to the observed track quality depreciations. To use the machine replacement problem, the cost of the machine renewal, i.e. the track update, should also be specified. Since there can be only one task scheduled at a time, the cost of executing a task should include the cost of not-executing the other tasks.

The state probabilities are obtained with the interacting multiple model (IMM), as described in \cite[Chapter 11]{barshalombook}. The mode-probabilities of IMM are associated with the state probabilities of the machine replacement problem. We assume that there are two motion models, both of which are constant velocity models having different process noise covariance matrices. The covariance matrix of two models are given as $\mathbf{Q}^2={100}^2\mathbf{Q}^1$, where $\mathbf{Q}^k$ denotes the covariance matrix for the $k$th model. The case of higher process noise covariance matrix refers to the case of a target in the stale state. The other case refers to the up-to-date state. We can say that the probability of target being in the up-to-date state is taken as the mode-probability of the model-$1$.

In addition, the track is considered good; when the trace of IMM mixed covariance matrix is within the allowed values. Otherwise, it is considered a bad track. Actions are UPD (update), similar to \textit{replace the machine} action comes with a cost $\cal K$ that will be explained later, and NUPD (not-update). In Table~\ref{tab:tsvsmp}, the analogy between the target selection and machine replacement problem is summarized. In the next section, we provide further details on the analogy given in Table~\ref{tab:tsvsmp}.

\stbl{\btb
\caption{Target Selection vs. Machine Replacement}\label{tab:tsvsmp}
\begin{tabular}{|l|c|c|}
	\hline
	Problem & Target Selection & Machine Replacement \\
	\hh
	\# of machines & $> 1$ & $1$\\
	\hline
	States & Up-to-date, & Good, \\
	&  Stale & Bad \\
	\hline
	Observations & good track, & conforming unit, \\
	& bad track & defective unit\\
	\hline
	Actions & update, not-update & replace, continue \\
	\hline
	Cost & target dependent & fixed \\
	\hline
\end{tabular}
\etb}

\subsection{Problem Model}

It is supposed that there are $\Nk$ targets at time $k$. The target selection problem emerges when there are more than one target concurrently requesting the track update among $\Nk$ targets. Then, the scheduler should decide which one of these targets is in need of track update more than others by using the information on the target states. There are $\Nk$ distinct Markov chains corresponding to each target, and the state transition probabilities of each target are assumed to be independent.

\subsubsection{Markov Chain Descriptions}
The target-$n$ obeys a $2$-state Markov chain with the following description:

\begin{itemize}
\item State is $x^n_k\in \{\nmv,\mv\}$, where $\nmv$ denotes the up-to-date state and $\mv$ denotes the stale state, with initial probability $P(x^n_0=i)=0.5$ for $i\in\{\nmv,\mv\}$.

\item Observation is $y^n_k\in \{gt,bt\}$, where $gt$ denotes a good track and $bt$ denotes a bad track.

\item Action is $u^n_k\in \{\text{NUPD},\text{UPD}\}$, where NUPD denotes the \textit{not-update} action and UPD denotes the \textit{update} action,
\end{itemize}
at time $k$ for $n=1,2,\dots,\Nk$.

\begin{figure}[!t]
	\centering
	\subfigure[Chain for the NUPD action.]{
		\includegraphics{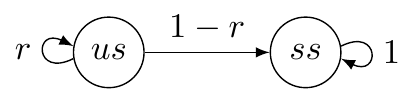}
		\label{fig:mcu0}
	}
	\subfigure[Chain for the UPD action.]{
		\includegraphics{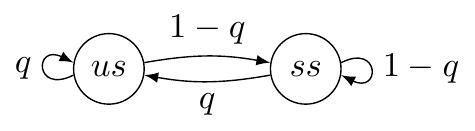}
		\label{fig:mcu1}
	}
	\caption{Markov chains for \subref{fig:mcu0} NUPD (not-update) and \subref{fig:mcu1} UPD (update) actions.}
	\label{fig:markovch}
\end{figure}

The state of a target is probabilistically evolving, such that if the target is in up-to-date state at time $k$, it will remain in that state with probability $r$ or it will change its state to the stale state with probability $1-r$ at time $k+1$. Once the target enters the stale state, it is assumed to remain in that state until the UPD action is taken, as shown in Fig.~\ref{fig:mcu0}. However, the UPD action may fail. A stale target moves to the up-to-date state with probability $q$ after taking the UPD action, as shown in Fig.~\ref{fig:mcu1}. If the NUPD action is taken, the transition of target states becomes a time-homogeneous Markov chain with the up-to-date state, a transient state, and the stale state, an absorbing state.

The conditional observation probabilities can be expressed as follows:
\bea
P(y^n_k=gt|x^n_k=\nmv)\eql \thtz,\\
P(y^n_k=gt|x^n_k=\mv)\eql \thto,\label{eq:theta1}\\
P(y^n_k=bt|x^n_k=\nmv)\eql 1-\thtz,\\
P(y^n_k=bt|x^n_k=\mv)\eql 1-\thto.
\eea

The expressions given above present the probability of having an observation matching the actual state of the target. More specifically, $\thtz$ denotes the probability of the measurement matching the state of the target, i.e. a measurement indicating a good track quality given that the target is in the up-to-date state.

\subsubsection{Cost Function}
Different from the classical machine replacement problem, the cost of updating a specific track (machine renewal) affects the cost of other tracks since there can be only one track that can be updated at an instant, and selecting a specific track for the update action leads to the track quality depreciation of other tracks. We propose to use the following cost function $\Knk$ taking into account the coupling of track quality scores for individual targets:
\begin{equation}
\Knk \triangleq \dfrac{\max \left\{m^{(0,\ell)}_k \xrq^\ell\right\}^{\Nk}_{\ell=1,\: \ell\neq n}}{m^{(1,n)}_k}\cdot \xrq^n,\label{eq:costpar}
\end{equation}
where $\xrq^n\in\{0,1\}$ indicates whether the target-$n$ requests a track update or not. The value of $m^{(1,n)}_k$ denotes the estimated improvement on the IMM mixed covariance of the target-$n$ by taking the UPD action $\big(\ukno\big)$. We assume that once a target is updated, the diagonal elements of the IMM mixed covariance matrix would be reduced due to the measurement update process. The value of $\max\big\{m^{(0,\ell)}_k\big\}^{\Nk}_{\ell=1,\: \ell\neq n}$ denotes the maximum of estimated deterioration on the IMM mixed covariance of all targets that are not updated. This set covers all targets except the target-$n$. From \eqref{eq:costpar}, it can be noted that the cost of updating target-$n$ is related with the cost of not-updating other targets.

The improvement or deterioration metric, $m^{(1,n)}_k$, $m^{(0,n)}_k$, can be taken as the trace of the IMM mixed covariance matrix. If the trace decreases at time $k+1$, an improvement on the track quality occurs; otherwise, the track quality deteriorates.

\begin{figure}[!t]
	\centering
	\includegraphics{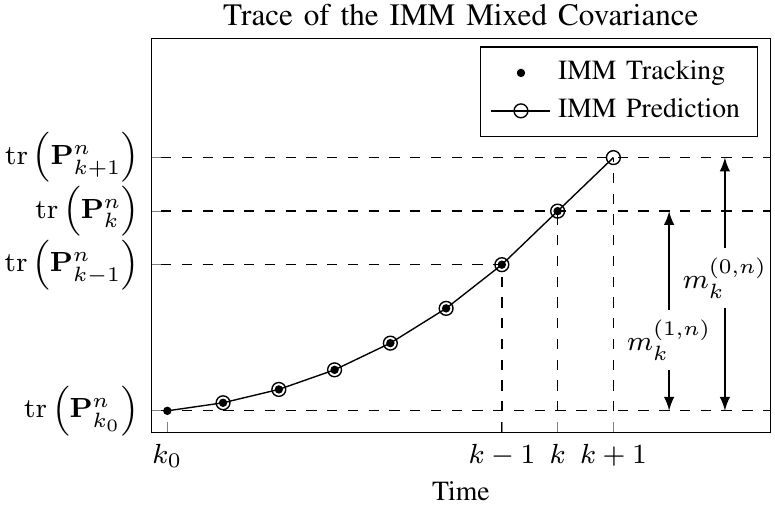}
	\caption{Description of the parameters used to find the cost function value.\label{fig:m0m1}}
\end{figure}

Fig.~\ref{fig:m0m1} is given to visualize the definitions for $m^{(0,n)}_k$ and $m^{(1,n)}_k$ are
\bea
m^{(0,n)}_k \eql \trace \Big(\mathbf{P}^n_{k+1}\Big) -\trace \Big(\mathbf{P}^n_{k_0}\Big),\label{eq:m0}\\
m^{(1,n)}_k \eql \trace \Big(\mathbf{P}^n_{k}\Big) -\trace \Big(\mathbf{P}^n_{k_0}\Big). \label{eq:m1}
\eea
Here, $\mathbf{P}^n_{k_0}$ is the IMM mixed covariance matrix of the target-$n$ at time $k_0$ when the target-$n$ has the latest measured track. We assume that if track is updated at next time $k+1$, the trace of the IMM mixed covariance matrix would be close to the trace of $\mathbf{P}^n_{k_0}$. Hence, $m^{(1,n)}_k$ and $m^{(0,n)}_k$ depend on $\mathbf{P}^n_{k_0}$.

In the next section, we derive the expressions required for the solution of target selection problem with the definition presented in this section.

\subsection{Derivation of Required Expressions}
The probability of being in the up-to-date state is
\begin{equation}
\infst=P(x^n_k=\nmv)
\end{equation}
for the target-$n$ at time $k$. Then, the probability of observing a good track is
\bea
\Pgt(\infst) \teql  P(y^n_k=gt)\nonu
\eql \sum\limits_{i\in\{\nmv,\mv\}} P(y^n_k=gt|x^n_k=i) P(x^n_k=i)\nonu
\eql (\thtz -\thto)\infst+\thto\label{eq:gtprob}.
\eea
Similarly, the probability of observing a bad track is
\bea
\Pbt(\infst) \teql  P(y^n_k=bt)\nonu
\eql \sum\limits_{i\in\{\nmv,\mv\}} P(y^n_k=bt|x^n_k=i) P(x^n_k=i)\nonu
\eql  1-(\thtz -\thto)\infst-\thto\label{eq:btprob}.
\eea
By applying the Bayes' theorem, the posterior probabilities for the up-to-date state are given as
\begin{IEEEeqnarray}{rCl}
	\IEEEeqnarraymulticol{3}{l}{
		P(x^n_k=\nmv|y^n_k=gt)
	}\nonu \quad\quad\quad\quad
	\eql \dfrac{P(y^n_k=gt|x^n_k=\nmv)P(x^n_k=\nmv)}{P(y^n_k=gt)}\nonu
	\eql \dfrac{\thtz \infst}{\Pgt(\infst)}\label{eq:postgt},\\
	\IEEEeqnarraymulticol{3}{l}{
		P(x^n_k=\nmv|y^n_k=bt)
	}\nonu \quad\quad\quad\quad \eql\dfrac{P(y^n_k=bt|x^n_k=\nmv)P(x^n_k=\nmv)}{P(y^n_k=bt)}\nonu
	\eql \dfrac{(1-\thtz) \infst}{\Pbt(\infst)}\label{eq:postbt}.
\end{IEEEeqnarray}
With the Markov property, it is assumed that $x^n_{k+1}$ is conditionally independent of $y^n_k$ \cite{kumarbook}, and hence, the conditional probability of the next state is $j$ given that $gt$ is observed and the NUPD action is taken in the current state is $i$ can be written as
\begin{IEEEeqnarray}{rCl}
\IEEEeqnarraymulticol{3}{l}{
P(x^n_{k+1}=j|x^n_k=i,y^n_k=gt,\uknz)
}\nonu \quad\quad
\eql P(x^n_{k+1}=j|x^n_k=i,\uknz)\label{eq:condind}\nonu
\eql P_{ij}(u^n_k),
\end{IEEEeqnarray}
where $i,j\in\{\nmv,\mv\}$. By using these expressions and the law of total probability, the conditional probabilities of the next state are obtained. The probability of being in the up-to-date state at next time given that a good track is observed and the NUPD action is taken at current time is expressed as
\begin{IEEEeqnarray*}{rCl}
\IEEEeqnarraymulticol{3}{l}{
	P(x^n_{k+1}=\nmv|y^n_k=gt,\uknz)\hspace{7.8em}}\\
\hspace{2em}\eql \sum\limits_{i\in\{\nmv,\mv\}}P(x^n_{k+1}=\nmv,x^n_k=i| \\
\IEEEeqnarraymulticol{3}{r}{ y^n_k=gt,\uknz)}\\
\eql\sum\limits_{i\in\{\nmv,\mv\}}P(x^n_{k+1}=\nmv|x^n_k=i,y^n_k=gt, \\
\IEEEeqnarraymulticol{3}{r}{\uknz)P(x^n_k=i|y^n_k=gt,\uknz)}\\
\eql \sum\limits_{i\in\{\nmv,\mv\},j=\nmv}P_{ij}(u^n_k)P(x^n_k=i|y^n_k=gt)\\
\eql r\cdot P(x^n_k=\nmv|y^n_k=gt)\\
\IEEEeqnarraymulticol{3}{r}{ +0\cdot P(x^n_k=\mv|y^n_k=gt)}\\
\eql \dfrac{r \thtz \infst}{\Pgt(\infst)}\IEEEyesnumber\label{eq:nextgt},
\end{IEEEeqnarray*}
and the probability of being in the up-to-date state at next time given that a bad track is observed and the NUPD action is taken at current time is
\begin{IEEEeqnarray*}{rCl}
\IEEEeqnarraymulticol{3}{l}{
P(x^n_{k+1}=\nmv|y^n_k=bt,\uknz)\hspace{7.8em}}\\
\hspace{2em}\eql \sum\limits_{i\in\{\nmv,\mv\}}P(x^n_{k+1}=\nmv,x^n_k=i| \\
\IEEEeqnarraymulticol{3}{r}{ y^n_k=bt,\uknz)}\\
\eql\sum\limits_{i\in\{\nmv,\mv\}}P(x^n_{k+1}=\nmv|x^n_k=i,y^n_k=bt, \\
\IEEEeqnarraymulticol{3}{r}{\uknz)P(x^n_k=i|y^n_k=bt,\uknz)}\\
\eql \sum\limits_{i\in\{\nmv,\mv\},j=\nmv}P_{ij}(u^n_k)P(x^n_k=i|y^n_k=bt)\\
\eql r\cdot P(x^n_k=\nmv|y^n_k=bt)\\
\IEEEeqnarraymulticol{3}{r}{ +0\cdot P(x^n_k=\mv|y^n_k=bt)}\\
\eql \dfrac{r(1-\thtz) \infst}{\Pbt(\infst)}\IEEEyesnumber\label{eq:nextbt}.
\end{IEEEeqnarray*}
Note that the conditional probabilities given in \eqref{eq:nextgt} and \eqref{eq:nextbt} can be considered the function of $\infst$, since $\thtz$ and $\thto$ are the global constants for the problem. We use the following notations for the conditional probabilities
\bea
\Hgt(\infst) \teql P(x^n_{k+1}=\nmv|y^n_k=gt,\uknz)\nonu
\eql \dfrac{r \thtz \infst}{(\thtz -\thto)\infst+\thto},\label{eq:hgt}\\
\Hbt(\infst) \teql P(x^n_{k+1}=\nmv|y^n_k=bt,\uknz)\nonu
\eql \dfrac{r(1-\thtz) \infst}{1-(\thtz -\thto)\infst-\thto}\label{eq:hbt}.
\eea

According to \cite[Lemma 1]{grosfeld06}, both $\Hgt(\infst)$ and $\Hbt(\infst)$ are continuous and strictly increasing functions for $0<\infst<1$, while $\Hgt(\infst)$ is strictly concave and $\Hbt(\infst)$ is strictly convex. Moreover, the inverse functions $\cHg^{-1}(\infst)$ of $\Hgt(\infst)$ and $\cHb^{-1}(\infst)$ of $\Hbt(\infst)$ exist, and they are strictly increasing for $0<\infst<r$, proofs can be found in \cite[Appendix A]{ocayirms}. The inverse function $\cHb^{-1}(\infst)$ is
\begin{equation}
\cHb^{-1}(\infst) =\dfrac{(1-\thto) \infst}{(\thtz -\thto)\infst+(1-\thtz) r}\label{eq:hbtint}
\end{equation}
for $0<\infst<r$. The existence of inverse leads to the use of function composition, i.e. $\Hbt\left(\cHb^{-1}(\infst)\right)=\infst$.

The functions $\Hgt(\infst)$ and $\Hbt(\infst)$ depend on the fixed parameters, $r$, $\thtz$ and $\thto$ as well. The critical point for choosing the fixed parameters is to ensure the criteria, $r\thtz>\thto$, so that $\Hgt(\mu^*)=\mu^*> 0$ does exist. The value of $\mu^*$ is computed as
\begin{equation}
\mu^*=\dfrac{r\thtz -\thto}{\thtz -\thto}.\label{eq:optmu}
\end{equation}

\begin{figure}[!t]
\centering
\subfigure[]{
\includegraphics{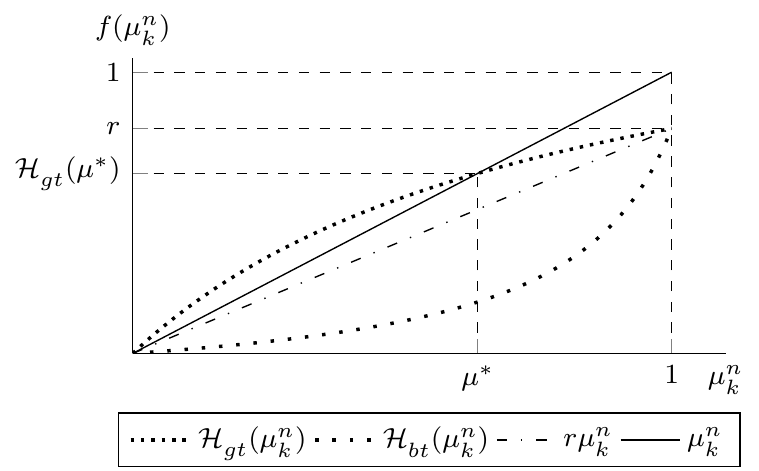}
    \label{fig:hplot}
}
\subfigure[]{
\includegraphics{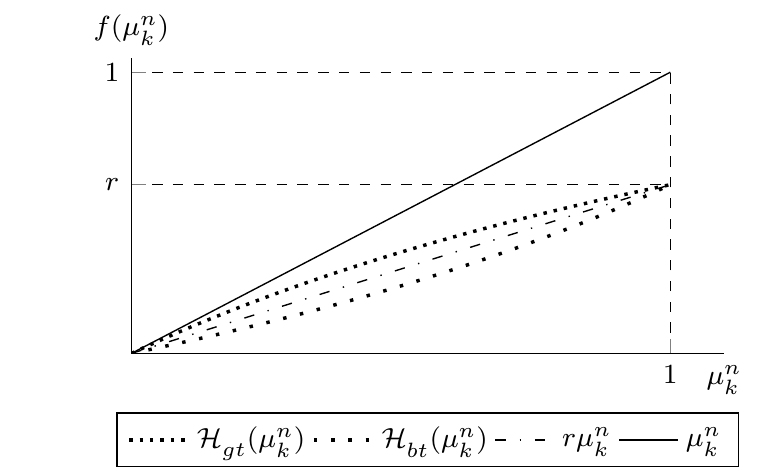}
    \label{fig:hplotr}
}
\caption{The functions $\Hgt(\infst)$ and $\Hbt(\infst)$ for \subref{fig:hplot} $r\thtz>\thto$, where $r=0.8$, $\thtz=0.9$, $\thto=0.4$, and \subref{fig:hplotr} $r\thtz<\thto$, where $r=0.6$, $\thtz=0.6$, $\thto=0.4$.}
\label{fig:hfplot}
\end{figure}

The point, $\mu^*$, divides the domain of $\infst$ into $2$ sub-domains, $\mu^*<\Hgt(\infst)<\infst$ for $\infst>\mu^*$ and $\Hgt(\infst)>\infst$ for $0<\infst<\mu^*$. If $r\thtz\leqslant\thto$, then $0<\Hgt(\infst)<\infst$ for $0<\infst\leqslant 1$, and hence, $\Hgt(\mu^*)=\mu^*$ does not exist, as shown in Fig.~\ref{fig:hfplot}. In Fig.~\ref{fig:hplot}, parameters satisfy $r\thtz>\thto$ and $\Hgt(\mu^*)=\mu^*$ exists. If the observation on the state is $gt$, one knows that the probability of being in the up-to-date state will increase at the next time step, i.e. $\infstn>\infst$, for $0<\infst<\mu^*$, whereas it will decrease for $\mu^*<\infst\leqslant 1$, as shown in Fig.~\ref{fig:hplot}. If $\mu^*$ does not exist, as shown in Fig.~\ref{fig:hplotr}, then the same probability always decreases, i.e. $\infstn<\infst$, irrespective of the observation. Thus, the provided observations can be considered to be informative, if $\mu^*$ exists, i.e. for $r\thtz>\thto$. The existence of $\mu^*$ is not critical for the implementation of the scheme, but important for its conceptual understanding.

Next, we discuss the value function and associated optimal policy for the infinite-horizon problem where the probability calculations given in this section is required in the computations.

\subsubsection{Infinite-Horizon Value Functions}
The solution to the stochastic control problem described is studied only for the infinite-horizon case. There is no fixed period which the problem restarts with a given set of initial parameters for the target tracking problem. For this problem, each target has a different motion characteristic, and hence, the target-based generalization of initial parameters is not practical. Furthermore, the infinite-horizon case solution has the advantage of requiring less computation. It only requires the computation of a fixed threshold for the asymptotic probability of being in the up-to-date state. Therefore, with a simple threshold policy, whenever the up-to-date state probability is less than the threshold, the optimal action becomes an update action.

The optimal action is determined via \textit{the optimal value function}, $\vf(\cdot)$, according to the probability of being in the up-to-date state, $\infst$,
\begin{equation}
\vf(\infst) = \max \left\{\vnupd(\infst),\vupd\right\},\label{eq:valfun}
\end{equation}
where $\vnupd(\infst)$ and $\vupd$ are the infinite-horizon value functions for NUPD and UPD actions, respectively as follows:
\begin{IEEEeqnarray}{rCl}
\vnupd(\infst)\eql \infst +\alpha\sum\limits_{i \in\{gt,bt\}}\clP_i(\infst)\vf\big(\clH_i(\infst)\big)\label{eq:valnupd}\IEEEeqnarraynumspace\nonu
\eql \infst +\alpha \Big[\Pgt(\infst)\vf\big(\Hgt(\infst)\big)\nonu
& &\hspace{4.5em} +\Pbt(\infst)\vf\big(\Hbt(\infst)\big)\Big],\label{eq:valnupd2}\\
\vupd \eql -\Knk+\vnupd(q),\label{eq:valupd}
\end{IEEEeqnarray}
where $\alpha$ is a discount factor satisfying $0< \alpha <1$.

The value functions \eqref{eq:valnupd} and \eqref{eq:valupd} depend explicitly on each other via $\vf(\infst)$ given in \eqref{eq:valfun}. Furthermore, $\vnupd(\infst)$, given in \eqref{eq:valnupd2}, is related to the optimal value function, $\vf(\cdot)$ through the functions $\Hgt(\infst)$ and $\Hbt(\infst)$, while $\vf(\cdot)$, given in \eqref{eq:valfun}, is already related to $\vnupd(\cdot)$.

It is not possible to express $\vnupd(\infst)$ without any assumptions on $\Hgt(\infst)$ and $\Hbt(\infst)$. Choosing $\thto=0$, as stated in \cite{grosfeld13}, is a practical choice which assumes that the probability of observing a good track given that corresponding target is in the stale state is $0$ and it also avoids always-decreasing probability of the up-to-date state by ensuring $r\thtz>\thto$, i.e. the existence of $\mu^*$. Then, it is only possible that a good track is observed from a target in the up-to-date state. With this assumption, \eqref{eq:valnupd} can be simplified as
\begin{IEEEeqnarray}{rCl}
\vnupd(\infst)\eql \infst +\alpha \big[\thtz \infst \vf(r)\IEEEeqnarraynumspace\nonu
& &\hspace{3.5em} +(1-\thtz \infst)\vf\big(\Hbt(\infst)\big)\big].\label{eq:valnupds}\IEEEeqnarraynumspace
\end{IEEEeqnarray}

Another simplification on the parameter $q$, the probability of transition to the up-to-date state upon the update action, as shown in Fig.~\ref{fig:mcu1}, is required. To make the model more realistic, perhaps pessimistic, we take $q=r$, ($r\neq 1$), meaning that update action may fail (due to radar sensing environment). Then, \eqref{eq:valupd} becomes
\begin{equation}
\vupd = -\Knk+\vnupd(r).\label{eq:valupd2}
\end{equation}

In order to evaluate the value of $\vnupd(r)$ from \eqref{eq:valnupds}, we need
\bea
\vf(r)\eql \max \left\{\vnupd(r),\vupd\right\}\nonu
\eql \max \left\{ \Knk+\vupd,\vupd\right\}.\label{eq:valfun2r}
\eea
Since $\Knk$ given in \eqref{eq:costpar} cannot be negative valued, the equation \eqref{eq:valfun2r} becomes
\begin{equation}
\vf(r)=\Knk+\vupd.\label{eq:valfunr}
\end{equation}
By using \eqref{eq:valfunr}, the simplified value function of NUPD action, \eqref{eq:valnupds} becomes
\begin{IEEEeqnarray}{rCl}
\vnupd(\infst) \eql  \infst +\alpha \big[\thtz \infst \left(\Knk+\vupd\right)\IEEEeqnarraynumspace\nonu
& &\hspace{3em} +(1-\thtz \infst)\vf\big(\Hbt(\infst)\big)\big].\label{eq:valnupdf}\IEEEeqnarraynumspace
\end{IEEEeqnarray}

\subsubsection{The Threshold Value for Decision-Making}

The threshold value for the target-$n$, which is denoted as $\thM$, is the solution of $\vnupd(\infst)=\vupd$, where LHS and RHS are given in \eqref{eq:valnupdf} and \eqref{eq:valupd2} respectively. Then, the decision-making process is given as
\begin{equation}
u^n_k=\begin{cases}
        \text{NUPD},& \infst\geqslant \thM\ \\
        \text{UPD},& \text{otherwise}\
      \end{cases}
\end{equation}
such that the action is no-update on the track if the threshold is exceeded.

\begin{figure}[!t]
\centering
\includegraphics{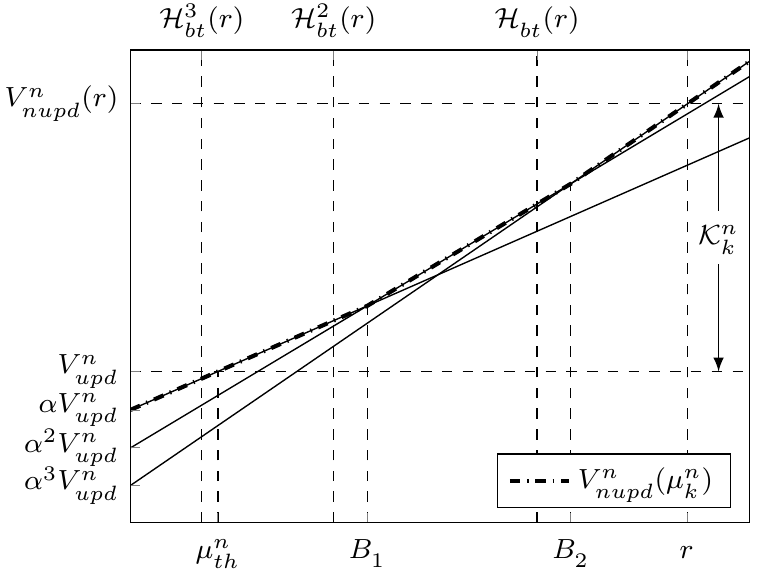}
\caption{The value function of NUPD action for $M=3$.}
\label{fig:seg}
\end{figure}

It is not straightforward to obtain \eqref{eq:valupd2} and \eqref{eq:valnupdf} owing to the presence of $\vnupd(r)$ in \eqref{eq:valupd2}. Fig.~\ref{fig:seg} is given to visualize the value function of NUPD action obtained from the basic parameters, $\alpha$, $\thtz$, $r$, $\Knk$. The function $\vnupd(\infst)$ is the pointwise maximum of linear functions cutting $\infst=0$ axis at $\alpha^3 \vupd$, $\alpha^2 \vupd$ and $\alpha \vupd$ for $0\leqslant\infst\leqslant 1$. Thus, it is a piecewise linear convex function, see \cite[Theorem 2]{grosfeld13}. The intersection points of these linear functions are the breakpoints of $\vnupd(\infst)$ such that $B_1=\cHb^{-1}(\thM)$ and $B_2=\cHb^{-1}(B_1)$. The number of linear functions, namely linear segments of $\vnupd(\infst)$, $M$, is determined with $\thM$ satisfying $r>\Hbt(r)>\cHb^2(r)>\cdots >\cHb^{M-2}(r)>\cHb^{M-1}(r)>\thM>\cHb^M(r)$, see \cite[Corollary 1]{grosfeld13}. To obtain $M$, the constraint $\cHb^M(r)<\thM$ is utilized, where $\cHb^M(r)$ is found by evaluating \eqref{eq:hbt} recursively. For each $M$ value starting from $M=1$, $\thM$ is computed and compared with $\cHb^M(r)$ until the constraint holds. The expression of $\thM$ depending on $M$ and the basic parameters will be provided later.

The relation between $\thM$ and $\vupd$ can be found from \eqref{eq:valnupdf} at $\infst=\thM$ by replacing $\vnupd(\thM)$ and $\vf\big(\Hbt(\thM)\big)$ with $\vupd$, where the latter is the result of $\Hbt(\thM)<\thM$, as shown in Fig.~\ref{fig:hplot}. Then, $\thM$ can be computed as
\begin{equation}
\thM = \dfrac{1-\alpha}{1+\alpha \thtz \Knk} \vupd,\label{eq:thres}
\end{equation}
if $\vupd$ is known, see \cite[Proposition 4]{grosfeld13}. Thus, the value of $\vupd$ is the most critical point of the calculations.

The expression of $\vupd$ depending on $M$ and the basic parameters is found by solving $M+1$ distinct equations, $\vf(r),\vf\big(\Hbt(r)\big),\dots,\vf\big(\cHb^{M-1}(r)\big)$ from \eqref{eq:valnupds} owing to $\vf(\infst)=\vnupd(\infst)$ for $\infst>\thM$, and $\vf\big(\cHb^M(r)\big)=\vupd$. Then, $\vupd$ can be computed by
\begin{equation}\label{eq:valupdf}
\vupd \triangleq \dfrac{\anM(1+\alpha \thtz  \Knk) -\Knk}{1-\alpha \thtz \anM-\bnM},
\end{equation}
where $\anM$ is defined as
\bea
\anM \teql \begin{cases}
        r,& M=1\ \\
        r+\displaystyle\sum\limits_{i=1}^{M-1}\alpha^i \cHb^i(r)&\\
        \hspace{3em}\prod\limits_{j=0}^{i-1} \big(1-\thtz \cHb^j(r) \big),& M\geqslant 2\
      \end{cases}\label{eq:anm}\IEEEeqnarraynumspace
\eea
and $\bnM$ is defined as
\begin{equation}
\bnM \triangleq \displaystyle\prod\limits_{i=0}^{M-1} \alpha \big(1-\thtz \cHb^i(r) \big),\label{eq:bnm}
\end{equation}
for $M\geqslant 1$ by supposing that $\vnupd(\infst)$ consists of $M$ segments, see \cite[Theorem 3 and 4]{grosfeld13}. Derivations of $\anM$ and $\bnM$ can be found in \cite[Chapter 4]{ocayirms}.

A careful examination of \eqref{eq:valupdf} reveals that $\vupd$ can be negative valued depending on the $\Knk$ value. If $\Knk$ is high enough, then $\vupd$ is negative valued, and hence, $\thM$ also becomes also negative according to \eqref{eq:thres}. Therefore, the NUPD action immediately becomes the optimal action for the negative threshold values since $\infst$ is always positive. This observation simplifies the computations for the decision-making process, since it is possibly to infer whether $\thM$ is negative or not, from $\Knk$. When
\begin{equation*}
\Knk>r/(1-\alpha r),
\end{equation*}
a degenerate policy emerges and the optimal action is always NUPD \cite[Proposition 2]{grosfeld08}, see the proof of Proposition 3 \cite[Chapter 4]{ocayirms}.

Then, the threshold value is determined by
\bea
\thM\eql \begin{cases}
        0,& \Knk>\dfrac{r}{1-\alpha r}\ \\[1em]
        \dfrac{1-\alpha}{1+\alpha \thtz \Knk}\vupd,& \text{otherwise.}\
      \end{cases}\label{eq:thr}
\eea

The computation of threshold $\thM$ is explicitly described in Table~\ref{algo:threshold}. Here, line 2 checks whether the parameters cause a degenerate policy or not. Lines 5 to 18 start with $M=1$ and increase $M$ until $\cHb^M(r)<\thM$ is satisfied, meanwhile, $\thM$ is computed by evaluating \eqref{eq:hbt}, \eqref{eq:anm}, \eqref{eq:bnm}, \eqref{eq:valupdf} and \eqref{eq:thr} respectively.

\renewcommand{\algorithmicrepeat}{\textbf{function} \textsc{Threshold}{$\big(\alpha,\:\thtz,\: r,\:\Knk\big)$}}
\stbl{\btb
\caption{Algorithm for Computing the Threshold Value}\label{algo:threshold}
\begin{tabular}{@{}c@{}}
\bmip
\begin{algorithmic}[1]
\REPEAT
\IF {$\Knk>r/(1-\alpha r)$}
	\STATE $\thM=0$
\ELSE
	\STATE $M = 1$
	\STATE compute $\cHb^M(r)$ from \eqref{eq:hbt}
	\STATE compute $\anM$  from \eqref{eq:anm}
	\STATE compute  $\bnM$ from \eqref{eq:bnm}
	\STATE compute $\vupd$ from \eqref{eq:valupdf}
	\STATE compute $\thM$ from \eqref{eq:thr}
	\WHILE {$\cHb^M(r)\geqslant\thM$}
		\STATE $M\!+\!+$
		\STATE compute $\cHb^M(r)$ from \eqref{eq:hbt}
		\STATE compute $\anM$  from \eqref{eq:anm}
		\STATE compute  $\bnM$ from \eqref{eq:bnm}
		\STATE compute $\vupd$ from \eqref{eq:valupdf}
		\STATE compute $\thM$ from \eqref{eq:thr}
	\ENDWHILE
\ENDIF
\RETURN $\thM$
\UNTIL
\end{algorithmic}
\emip
\end{tabular}
\etb}

We present the outputs of the algorithm in Table~\ref{algo:threshold} for some special cases in Table~\ref{tab:kval}. This table can also be used for debugging purposes. Here, we assume that $\alpha=0.99$ and other basic parameters, $\thtz$, $r$ and $\Knk$ are changed to obtain distinct infinite-horizon value functions. The algorithm outputs, namely the threshold values and the numbers of segments, are given in Table~\ref{tab:kval}.

Some comments on the data of Table~\ref{tab:kval} can be given as follows:

\begin{itemize}
\item The higher $r$ makes $\thM$ higher since $\anM$ increases with $r$, and $\vnupd$ also increases. This statement can be deduced from \eqref{eq:valupdf}.
\item The higher $\Knk$ makes $\thM$ smaller. That is
\[\Knk< r/(1-\alpha r) \wedge \Knk\rightarrow r/(1-\alpha r)\implies \thM\rightarrow 0.\]
\item $M$ depends on both $\thtz$ and $\Knk$, as illustrated more clearly for $\Knk=2.5$ in Table~\ref{tab:kval}.
\end{itemize}

\stbl{\btb
\caption{Comparison of the Threshold Value and Number of Segments for Different $\Knk$ Values with $\alpha=0.99$}
\label{tab:kval}
\begin{tabular}{|c||l|c|c|c||c|c|c|}
\hline
& & \multicolumn{3}{c||}{$r=0.90$} & \multicolumn{3}{c|}{$r=0.95$}\\
\cline{2-8}
$\Knk$& $\thtz$& $0.75$ & $0.80$ & $0.90$ & $0.75$ & $0.80$ & $0.90$ \\
\hh
$0.1$& $\thM$ & $0.8069$ & $0.8073$ & $0.8082$ & $0.8569$ & $0.8573$ & $0.8582$ \\
\cline{2-8}
& $M$ & $1$ & $1$ & $1$ & $1$ & $1$ & $1$  \\
\hh
$1.0$& $\thM$ & $0.3984$ & $0.3933$ & $0.3799$ & $0.4667$ & $0.4611$ & $0.4464$ \\
\cline{2-8}
& $M$ & $2$ & $2$ & $2$ & $2$ & $2$ & $2$  \\
\hh
$2.5$& $\thM$ & $0.1809$ & $0.1755$ & $0.1730$ & $0.2503$ & $0.2429$ & $0.2315$ \\
\cline{2-8}
& $M$ & $3$ & $3$ & $2$ & $3$ & $3$ & $2$  \\
\hline
\end{tabular}
\etb}

Up to now, we have discussed how to decide whether a \textit{single target} requires an update action or not based on its estimated track accuracy and calculated threshold given by the suggested algorithm. For the multiple target tracking scenarios, there can be a situation that several targets requiring an update at the same time, i.e. the up-to-date state probability falls below the corresponding threshold level. For such cases, we need to develop a policy for the selection of the most suitable target.

\subsection{Problem Solution: Target Selection Policies}
We present three policies for the selection of the track to be updated. The first policy is based on the machine replacement problem and uses the threshold policy for the selection of track. The other policies use IMM outputs, but not the threshold value; hence, they are simpler to implement.

\subsubsection{Decision Policy (DecP)} This policy selects the track to be updated among the tracks satisfying the condition $\infst<\thM$ for $n=\{1,2, \ldots, N_k\}$. It should be remembered that this condition is analogous to the decision of machine replacement.

The decision policy (DecP) selects the target-$i$ according to
\bea
i \eql \underset{n\in\{1,2,\dots,\Nk\}}{\operatorname{argmin}} \left\{\infst-\thM\right\}.\label{eq:besttar}\\
&&\sbjctb\sbjct \: \xrq^n=1\nonu
&&\sbjcts \: \infst<\thM\nonumber
\eea
Ideally, the condition $\infst<\thM$ should be satisfied only if the track is sufficiently degraded. If none of the tracks is not sufficiently degraded, there is no track update. Hence, under the ideal conditions, the radar resources are not wasted by updating the tracks solely by the lateness value.

It should be remembered that this method has multiple criterion to be satisfied to grant a track update. First, by the requirement of the TB method, $\tmb$ value should be non-negative; hence, the lateness parameter should be either zero or a positive number. Second, the track should be sufficiently degraded, which is a condition checked by $\infst<\thM$. Third, among all tracks satisfying first two conditions, the track with the highest gap to threshold is selected.

\subsubsection{Tracking Error Minimization Policy (MinTE)}
This is a greedy policy implementing the update of the track based on the IMM mixed covariance matrices of the targets. The aim is to select the target among the set of targets with non-negative $\tmb$ value and the worst case tracking error. To select the target-$i$, this method takes into account the mixed covariance matrix, but not the mode-probabilities of IMM:
\bea
i \eql \underset{n\in\{1,2,\dots,\Nk\}}{\operatorname{argmax}} \left\{\trace \Big(\mathbf{P}^n_{k}\Big)\right\}.\label{eq:minimerr}\\
&&\sbjctb\sbjct \: \xrq^n=1\nonumber
\eea

It should be noted that the target with the highest trace of IMM mixed covariance matrix may not necessarily correspond to a rapidly maneuvering target. This policy does not exert any effort in detecting the maneuvering action of the target.

\subsubsection{Pursuing the Most Maneuvering Target Policy (PurMM)}
It should be remembered that the probability of up-to-date state is given as $\infst$. Then, the updated target can be chosen according to the product of the trace of IMM mixed covariance matrix and the stale state probability, $1-\infst$:
\bea
i\eql\underset{n\in\{1,2,\dots,\Nk\}}{\operatorname{argmax}} \left\{\left(1-\infst \right)\trace \Big(\mathbf{P}^n_{k}\Big)\right\}. \label{eq:pursuit}\\
&&\sbjctb\sbjct \: \xrq^n=1\nonumber
\eea
This method aims to give higher priority to the targets with having a high probability of maneuvering, namely probability of the mode with high process noise. Hence, this method is called as the method of pursuing the most maneuvering target (PurMM).

\section{Numerical Comparisons}
\label{sec:expres}

The assumed instrumented range for the simulator is $200$ km. Within this detection range, the assigned priority changes from $1$ to $5$ according to the detected target range with a range step of $40$ km. For example, the target at the range of $60$ km is assigned to the priority level $4$, which is the second highest priority level. The measurement noise is ${\cal{N}}(0,\sigma_r^2)$ and ${\cal{N}}(0,\sigma_a^2)$ for range and azimuth, respectively, where $\sigma_r=80$ m and $\sigma_a=3$ mrad. Further details on radar simulator can be found in \cite{ocayirms}. Due to the nature of conventional TB scheduler, the target selection policies are applied on the targets with the same priority. Hence, the tracking error of targets at only similar ranges are compared by these policies.

\subsection{Case 1: Single Target Tracking Case} To illustrate the improvement brought by the DecP, we compare the performance of TB method utilizing the DecP with the conventional TB method. The conventional TB method does not utilize the track information provided by IMM. Hence, its performance is expected to be inferior to the one utilizing the DecP. Our goal is to contrast the difference between the two.

\begin{figure}[!t]
\centering
\subfigure[The conventional TB method.]{
\includegraphics{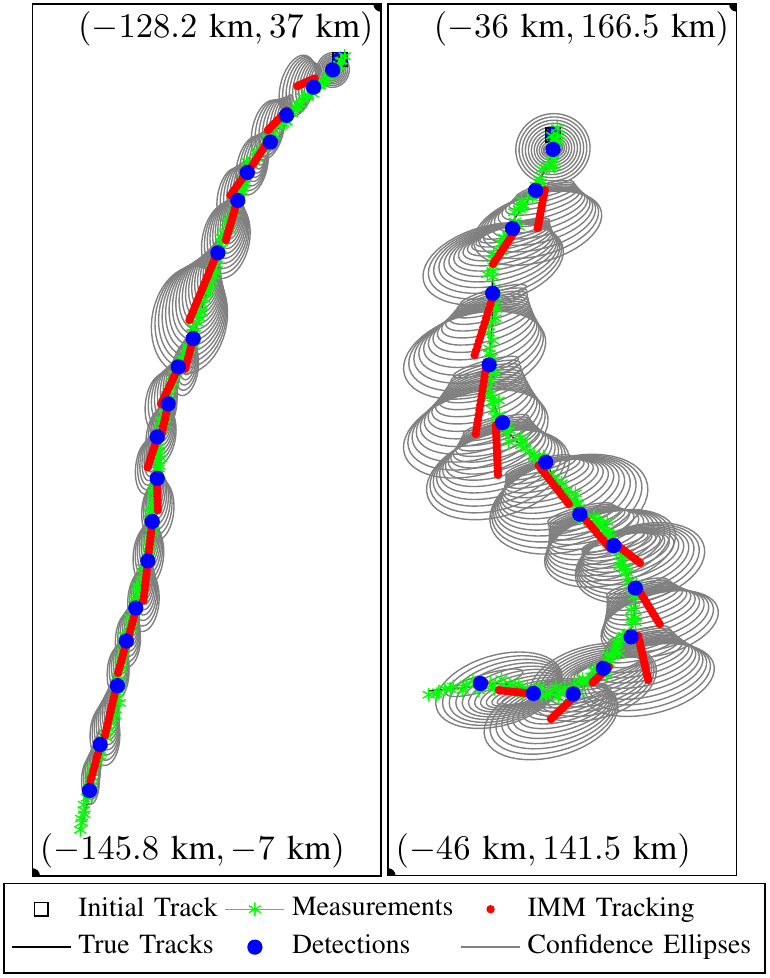}
    \label{fig:trgtcftul}
}
\subfigure[The TB method with DecP.]{
\includegraphics{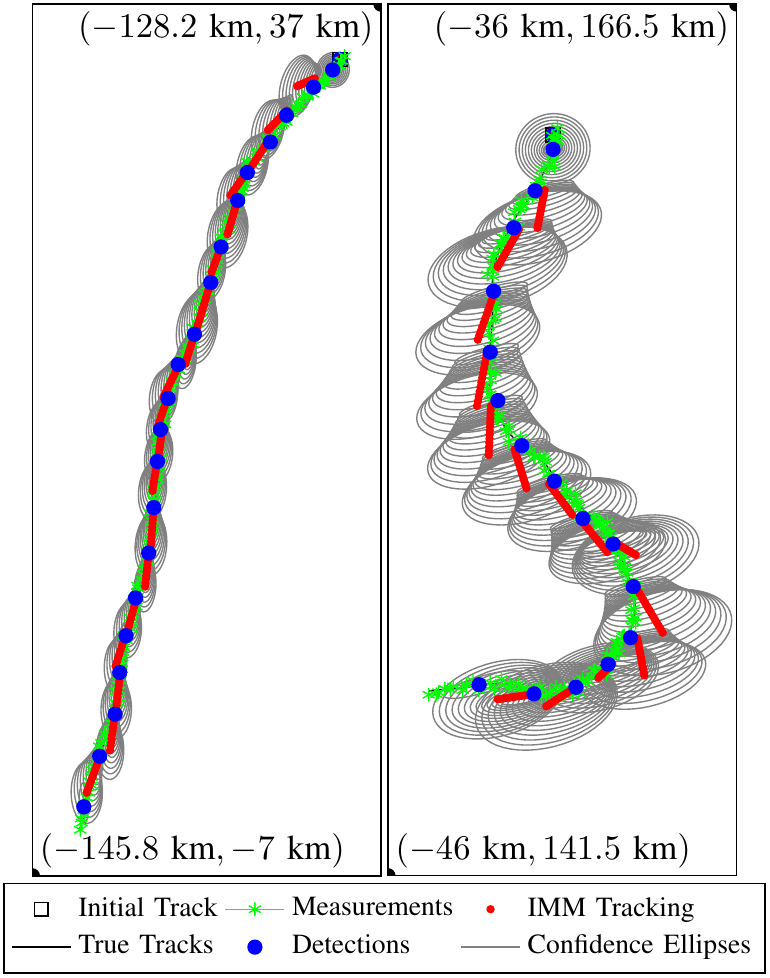}
    \label{fig:trgtdecp}
}
\caption{Tracking of non-maneuvering (left) and maneuvering (right) targets by using \subref{fig:trgtcftul} conventional TB method and \subref{fig:trgtdecp} TB method with DecP.}
\label{fig:mnvrng}
\end{figure}

The Fig.~\ref{fig:trgtcftul} shows the performance of conventional method on non-maneuvering (left panel of Fig.~\ref{fig:trgtcftul}) and maneuvering (right panel) targets. The tracking performance of the TB method with DecP is given in Fig.~\ref{fig:trgtdecp}. In this figure, the confidence ellipses are drawn to illustrate the tracking performance. A visual comparison of top and bottom panels of Fig.~\ref{fig:mnvrng} immediately reveals that the DecP yields a better performance by refraining from, or postponing, unnecessary track updates.

\begin{figure}[!t]
\centering
\subfigure[The conventional TB method.]{
\includegraphics{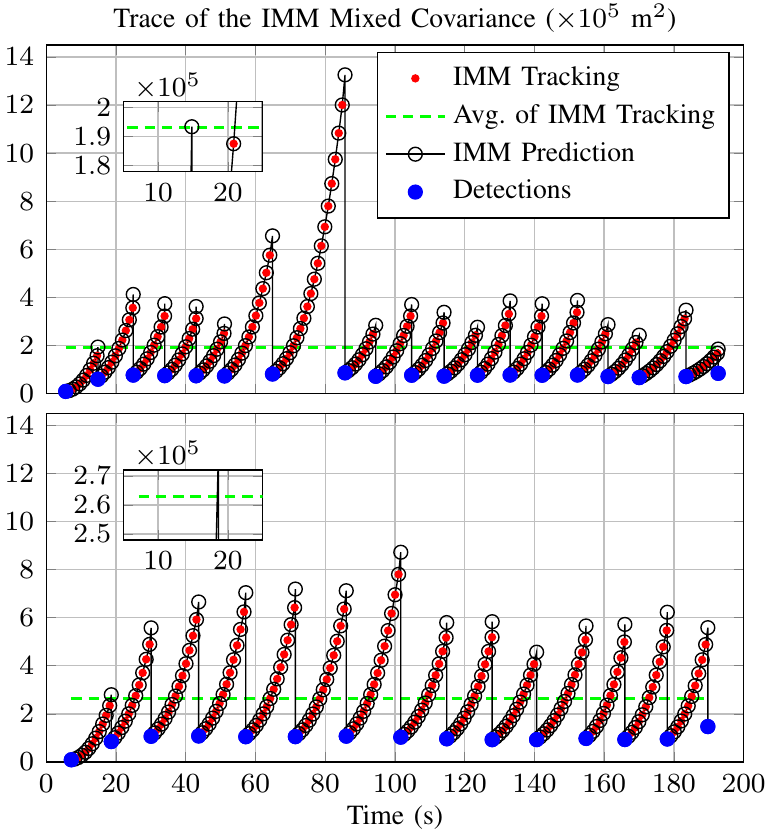}
    \label{fig:errcftul}
}
\subfigure[The TB method with DecP.]{
\includegraphics{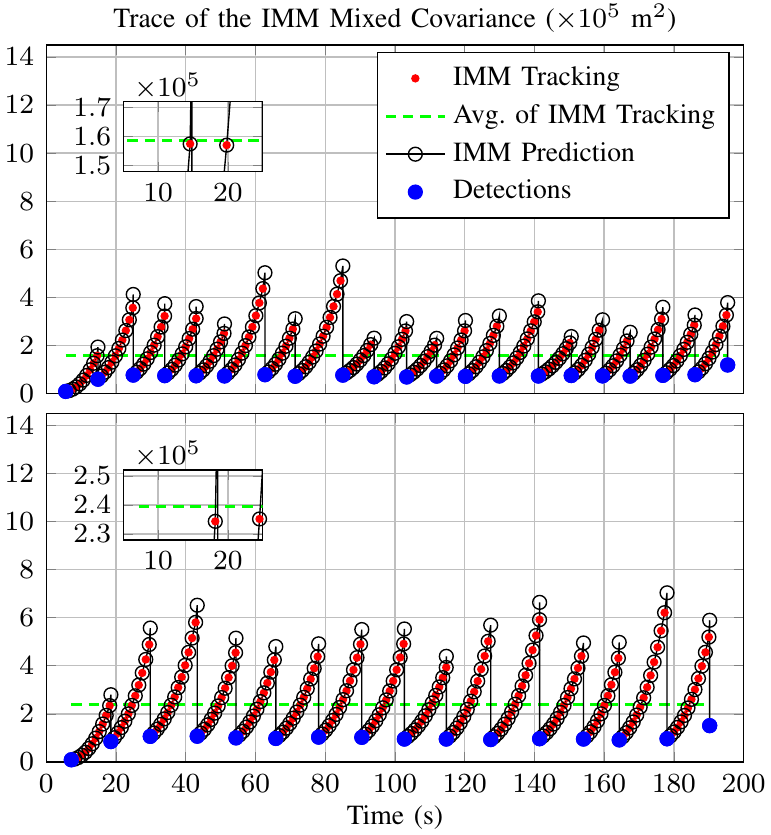}
    \label{fig:errdecp}
}
\caption{Trace of IMM mixed covariance matrices for non-maneuvering (top) and maneuvering (bottom) targets by using \subref{fig:errcftul} conventional TB and \subref{fig:errdecp} TB method with DecP.}
\label{fig:trackerr}
\end{figure}

More specifically, for the non-maneuvering target, the average tracking error, namely the average of the trace of IMM mixed covariance matrices, decreases from $1.93\kxx^5$ m$^2$, given in the top part of Fig.~\ref{fig:errcftul}, to $1.59\kxx^5$ m$^2$, given in the top part of Fig.~\ref{fig:errdecp}. For the maneuvering target, the average tracking error decreases from $2.63\kxx^5$ m$^2$ to $ 2.39\kxx^5$ m$^2$. Furthermore, the maximum value of the tracking error is also smaller with the DecP, which is a criterion that can be especially important for rapidly maneuvering targets.

However, it is important to remind that the DecP does not guarantee a better operation at every run, but can present significant improvements in the scenarios where the beginning of target maneuvering can be effectively sensed with the IMM mode-probabilities.

\def\theaddecnf{&  Conv.   &  DecP   &  Neural N.   &  Fuzzy L.  }
\def\theaddec{&  Conv.   &  DecP   &  MinTE   &  PurMM  }
\def\disthead{& \multicolumn{4}{c|}{Distributions of standings in avg. of errors}}
\def\clmnnt{$\#$ of tracking tasks}
\def\clmnns{$\#$ of surveillances}
\def\clmnnpd{$\#$ of prob. drops}
\def\clmnocc{Occupancy ($\%$)}
\def\clmncst{Cost (s$^2$)}
\def\clmnaoe{Avg. of errors (m$^2$)}
\def\clmnfr{Best}
\def\clmnsr{Runner-up}
\def\clmntr{Honorable Mention}
\def\clmnlr{Last}

\subsection{Case 2: Multiple Target Tracking Case} The proposed methods are evaluated for the scenario of multiple targets in addition to the surveillance tasks. The target tracks are randomly generated for each scenario of $200$ seconds. Each target has randomly chosen transition probability matrix out of five matrices, while the IMM tracker makes use of fixed transition probabilities for all tracks. There are $100$ distinct scenarios for each comparison case. The comparisons are made on the average of the scheduler performance that is measured with the following criteria:

\begin{itemize}
\item \textit{The number of probable drops} is the number of updates that are too late for target tracking. The probable drop occurs when the update interval exceeds the sum of task update time and allowable lateness.
\item \textit{Cost} is the sum of weighted lateness values squares after each scheduling epochs. The priority values are assigned as the weights.
\item \textit{Average of errors} is the average of the trace of IMM mixed covariance matrices of all targets.
\item \textit{Occupancy} is the ratio of utilized radar time to the total available time interval.
\end{itemize}

In Table~\ref{tab:compdec15} and \ref{tab:compdec25}, the suggested methods are compared for the scenarios of $15$ and $25$ in-track-targets, respectively. Our main goal is to compare the performance statistics resulting from the application of the conventional TB method and TB method augmented with the suggested policies.

In order to illustrate the effect of the loading condition more explicitly, we assume that the radar system can utilize multiple-frequency bands concurrently. An increase in the number of frequency bands reduces the load seen from the resource management side \cite{ocayirms}.
The number of frequency bands is selected as $2$ in Table~\ref{tab:compdec152} and \ref{tab:compdec252}, and the number of bands is selected as $7$ in Table~\ref{tab:compdec157} and \ref{tab:compdec257}.

When the proposed methods, (DecP, MinTE and PurMM) are compared with the conventional TB method, it can be seen that the DecP is the method which has the smallest number of probable drops and the smallest average of errors for all cases. Yet, the alternatives to the DecP (MinTE and PurMM) are almost equally good for this case.

\stbl{\btb
\caption{Comparison of the Decision Policies for $15$ Targets}
\label{tab:compdec15}
\subfigure[The number of frequency bands is $2$.]{\label{tab:compdec152}
\begin{tabular}{|L{0.28\columnwidth}|C{0.169\columnwidth}|C{0.169\columnwidth}|C{0.169\columnwidth}|C{0.169\columnwidth}|}
\hline
& \multicolumn{4}{c|}{Average of statistics after $100$ simulations}\\
\cline{2-5}
\theaddec \\
\hh
\clmnnt & $288.91$ & $\mathbf{292.11}$ & $289.73$ & $289.20$\\
\clmnns & $\mathbf{17.52}$ & $17.43$ & $17.37$ & $\mathbf{17.52}$\\
\clmnnpd & $23.76$ & $\mathbf{23.04}$ & $23.12$ & $23.08$\\
\clmnocc & $48.43$ & $\mathbf{48.69}$ & $48.37$ & $48.47$\\
\clmncst & $\mathbf{9.13\kx^4}$ & $1.01\kx^5$ & $9.49\kx^4$ & $1.11\kx^5$\\
\clmnaoe & $4.70\kx^5$ & $\mathbf{3.69\kx^5}$ & $3.97\kx^5$ & $4.64\kx^5$\\
\hh
\disthead\\
\hh
\clmnfr & $20$ & $\mathbf{38}$ & $19$ & $23$\\
\clmnsr & $24$ & $24$ & $\mathbf{33}$ & $19$\\
\clmntr & $25$ & $21$ & $\mathbf{30}$ & $24$\\
\clmnlr & $31$ & $17$ & $18$ & $\mathbf{34}$\\
\hline
\end{tabular}}

\subfigure[The number of frequency bands is $7$.]{\label{tab:compdec157}
\begin{tabular}{|L{0.28\columnwidth}|C{0.169\columnwidth}|C{0.169\columnwidth}|C{0.169\columnwidth}|C{0.169\columnwidth}|}
\hline
& \multicolumn{4}{c|}{Average of statistics after $100$ simulations}\\
\cline{2-5}
\theaddec \\
\hh
\clmnnt & $502.54$ & $\mathbf{504.67}$ & $504.22$ & $503.18$\\
\clmnns & $\mathbf{19.13}$ & $19.10$ & $18.98$ & $19.09$\\
\clmnnpd & $8.38$ & $\mathbf{8.22}$ & $8.39$ & $8.30$\\
\clmnocc & $74.17$ & $\mathbf{74.32}$ & $74.18$ & $74.15$\\
\clmncst & $9.49\kx^2$ & $8.13\kx^2$ & $\mathbf{7.28\kx^2}$ & $9.58\kx^2$\\
\clmnaoe & $8.95\kx^4$ & $\mathbf{8.77\kx^4}$ & $8.82\kx^4$ & $8.89\kx^4$\\
\hh
\disthead\\
\hh
\clmnfr & $20$ & $26$ & $\mathbf{32}$ & $22$\\
\clmnsr & $24$ & $\mathbf{30}$ & $17$ & $29$\\
\clmntr & $16$ & $\mathbf{35}$ & $26$ & $23$\\
\clmnlr & $\mathbf{40}$ & $9$ & $25$ & $26$\\
\hline
\end{tabular}}
\etb}

Tables also illustrate the performance comparison of the methods in a competitive sense. From the bottom part of the tables, it can be noted the conventional method has most frequently provided the poorest tracking error performance for the duration of complete scenarios, while the DecP has the smallest number of bad performances.

In Table~\ref{tab:compranks}, the distributions of standings given in Table~\ref{tab:compdec15} and \ref{tab:compdec25} are combined for a clearer comparison. Table~\ref{tab:compranks} indicates that the DecP is the most frequently successful policy among the four. It can be said that the DecP successfully traded the unnecessary track updates of targets having accurately predictable tracks, e.g. non-maneuvering targets, with the track quality depreciating targets, e.g. maneuvering targets. This conclusion can be further justified by examining the average of errors criteria in the tables where the DecP is the best policy in all cases. Hence, as in the single target case, the DecP, in essence, manages to ``detect'' the beginning of a maneuver successfully and does not grant unnecessary updates to a track in spite of its potentially large lateness value. Interested readers may examine \cite{ocayirms} for more comparisons.


\stbl{\btb
\caption{Comparison of the Decision Policies for $25$ Targets}
\label{tab:compdec25}
\subfigure[The number of frequency bands is $2$.]{\label{tab:compdec252}
\begin{tabular}{|L{0.28\columnwidth}|C{0.169\columnwidth}|C{0.169\columnwidth}|C{0.169\columnwidth}|C{0.169\columnwidth}|}
\hline
& \multicolumn{4}{c|}{Average of statistics after $100$ simulations}\\
\cline{2-5}
\theaddec \\
\hh
\clmnnt & $294.93$ & $296.39$ & $296.66$ & $\mathbf{298.07}$\\
\clmnns & $24.55$ & $24.55$ & $24.40$ & $\mathbf{24.63}$\\
\clmnnpd & $38.40$ & $\mathbf{37.37}$ & $37.98$ & $37.94$\\
\clmnocc & $55.70$ & $55.84$ & $55.72$ & $\mathbf{56.11}$\\
\clmncst & $5.21\kx^5$ & $5.22\kx^5$ & $\mathbf{4.79\kx^5}$ & $5.28\kx^5$\\
\clmnaoe & $8.73\kx^5$ & $\mathbf{7.44\kx^5}$ & $8.60\kx^5$ & $8.14\kx^5$\\
\hh
\disthead\\
\hh
\clmnfr & $20$ & $28$ & $18$ & $\mathbf{34}$\\
\clmnsr & $26$ & $\mathbf{31}$ & $21$ & $22$\\
\clmntr & $25$ & $24$ & $\mathbf{32}$ & $19$\\
\clmnlr & $\mathbf{29}$ & $17$ & $\mathbf{29}$ & $25$\\
\hline
\end{tabular}}

\subfigure[The number of frequency bands is $7$.]{\label{tab:compdec257}
\begin{tabular}{|L{0.28\columnwidth}|C{0.169\columnwidth}|C{0.169\columnwidth}|C{0.169\columnwidth}|C{0.169\columnwidth}|}
\hline
& \multicolumn{4}{c|}{Average of statistics after $100$ simulations}\\
\cline{2-5}
\theaddec \\
\hh
\clmnnt & $538.33$ & $\mathbf{538.74}$ & $537.18$ & $537.53$\\
\clmnns & $25.86$ & $25.94$ & $26.02$ & $\mathbf{26.10}$\\
\clmnnpd & $39.48$ & $\mathbf{39.33}$ & $40.48$ & $40.10$\\
\clmnocc & $83.76$ & $83.88$ & $83.79$ & $\mathbf{83.92}$\\
\clmncst & $\mathbf{1.68\kx^5}$ & $1.89\kx^5$ & $1.80\kx^5$ & $1.89\kx^5$\\
\clmnaoe & $3.35\kx^5$ & $\mathbf{3.28\kx^5}$ & $3.49\kx^5$ & $3.58\kx^5$\\
\hh
\disthead\\
\hh
\clmnfr & $22$ & $\mathbf{35}$ & $21$ & $22$\\
\clmnsr & $\mathbf{35}$ & $25$ & $21$ & $19$\\
\clmntr & $21$ & $28$ & $16$ & $\mathbf{35}$\\
\clmnlr & $22$ & $12$ & $\mathbf{42}$ & $24$\\
\hline
\end{tabular}}
\etb}

\stbl{\btb
\caption{Overall Distributions of Standings in Average of Errors After $400$ Simulations}
\label{tab:compranks}
\begin{tabular}{|L{0.28\columnwidth}|C{0.169\columnwidth}|C{0.169\columnwidth}|C{0.169\columnwidth}|C{0.169\columnwidth}|}
\hline
\theaddec \\
\hh
\clmnfr & $82$ & $\mathbf{127}$ & $90$ & $101$\\
\clmnsr & $109$ & $\mathbf{110}$ & $92$ & $89$\\
\clmntr & $87$ & $\mathbf{108}$ & $104$ & $101$\\
\clmnlr & $\mathbf{122}$ & $55$ & $114$ & $109$\\
\hline
\end{tabular}
\etb}

\subsection{Comparisons with Task Prioritization Methods} We compare the conventional TB and DecP with two other task prioritization methods based on neural network \cite{komorniczak,komorniczak2} and fuzzy logic \cite{miranda}. The detailed descriptions on these methods such as the choice of training set for the neural network based scheme and the membership functions for the fuzzy logic based scheme can be found in Appendix. Similar to the decision policies, these methods incorporate the tracking error into decision-making. In addition, they use some other inputs such as the radial velocity and the allowable lateness for the task prioritization. Unlike the decision policies, which are applied only when the conventional TB requires selecting one of targets having the same priority level, the task prioritization methods are continually applied.

In Table~\ref{tab:compdecnf15} and \ref{tab:compdecnf25}, the DecP is compared with the task prioritization methods for $15$ and $25$ targets, respectively. From the viewpoint of minimum average tracking error, the DecP policy remains as the best choice. On the other hand, the task prioritization methods provides the minimum number of probable drops due to the inclusion of the allowable lateness parameter in the scheduler design.

\stbl{\btb
\caption{Comparison with Task Prioritization Methods for $15$ Targets}
\label{tab:compdecnf15}
\subfigure[The number of frequency bands is $2$.]{\label{tab:compdecnf152}
\begin{tabular}{|L{0.28\columnwidth}|C{0.169\columnwidth}|C{0.169\columnwidth}|C{0.169\columnwidth}|C{0.169\columnwidth}|}
\hline
& \multicolumn{4}{c|}{Average of statistics after $100$ simulations}\\
\cline{2-5}
\theaddecnf \\
\hh
\clmnnt & $288.91$ & $\mathbf{292.11}$ & $287.21$ & $283.26$\\
\clmnns & $\mathbf{17.52}$ & $17.43$ & $16.70$ & $16.61$\\
\clmnnpd & $23.76$ & $23.04$ & $21.75$ & $\mathbf{21.34}$\\
\clmnocc & $48.43$ & $\mathbf{48.69}$ & $47.46$ & $47.16$\\
\clmncst & $\mathbf{9.13\kx^4}$ & $1.01\kx^5$ & $1.29\kx^5$ & $1.56\kx^5$\\
\clmnaoe & $4.70\kx^5$ & $\mathbf{3.69\kx^5}$ & $5.13\kx^5$ & $6.94\kx^5$\\
\hh
\disthead\\
\hh
\clmnfr & $21$ & $\mathbf{49}$ & $24$ & $6$\\
\clmnsr & $\mathbf{37}$ & $25$ & $22$ & $16$\\
\clmntr & $22$ & $19$ & $\mathbf{35}$ & $24$\\
\clmnlr & $20$ & $7$ & $19$ & $\mathbf{54}$\\
\hline
\end{tabular}}

\subfigure[The number of frequency bands is $7$.]{\label{tab:compdecnf157}
\begin{tabular}{|L{0.28\columnwidth}|C{0.169\columnwidth}|C{0.169\columnwidth}|C{0.169\columnwidth}|C{0.169\columnwidth}|}
\hline
& \multicolumn{4}{c|}{Average of statistics after $100$ simulations}\\
\cline{2-5}
\theaddecnf \\
\hh
\clmnnt & $502.54$ & $504.67$ & $\mathbf{509.66}$ & $509.08$\\
\clmnns & $\mathbf{19.13}$ & $19.10$ & $18.56$ & $18.48$\\
\clmnnpd & $8.38$ & $8.22$ & $\mathbf{7.45}$ & $7.58$\\
\clmnocc & $74.17$ & $74.32$ & $\mathbf{74.38}$ & $74.22$\\
\clmncst & $9.49\kx^2$ & $8.13\kx^2$ & $\mathbf{7.90\kx^2}$ & $4.95\kx^3$\\
\clmnaoe & $8.95\kx^4$ & $\mathbf{8.77\kx^4}$ & $8.88\kx^4$ & $9.49\kx^4$\\
\hh
\disthead\\
\hh
\clmnfr & $15$ & $\mathbf{36}$ & $32$ & $17$\\
\clmnsr & $\mathbf{31}$ & $28$ & $23$ & $18$\\
\clmntr & $\mathbf{32}$ & $27$ & $21$ & $20$\\
\clmnlr & $22$ & $9$ & $24$ & $\mathbf{45}$\\
\hline
\end{tabular}}
\etb}


\stbl{\btb
\caption{Comparison with Task Prioritization Methods for $25$ Targets}
\label{tab:compdecnf25}
\subfigure[The number of frequency bands is $2$.]{\label{tab:compdecnf252}
\begin{tabular}{|L{0.28\columnwidth}|C{0.169\columnwidth}|C{0.169\columnwidth}|C{0.169\columnwidth}|C{0.169\columnwidth}|}
\hline
& \multicolumn{4}{c|}{Average of statistics after $100$ simulations}\\
\cline{2-5}
\theaddecnf \\
\hh
\clmnnt & $294.93$ & $\mathbf{296.39}$ & $290.72$ & $276.92$\\
\clmnns & $\mathbf{24.55}$ & $\mathbf{24.55}$ & $23.84$ & $23.62$\\
\clmnnpd & $38.40$ & $37.37$ & $\mathbf{35.03}$ & $35.99$\\
\clmnocc & $55.70$ & $\mathbf{55.84}$ & $54.58$ & $53.16$\\
\clmncst & $\mathbf{5.21\kx^5}$ & $5.22\kx^5$ & $5.76\kx^5$ & $7.10\kx^5$\\
\clmnaoe & $8.73\kx^5$ & $\mathbf{7.44\kx^5}$ & $9.23\kx^5$ & $1.36\kx^6$\\
\hh
\disthead\\
\hh
\clmnfr & $24$ & $\mathbf{38}$ & $30$ & $8$\\
\clmnsr & $\mathbf{39}$ & $36$ & $16$ & $9$\\
\clmntr & $25$ & $14$ & $\mathbf{35}$ & $26$\\
\clmnlr & $12$ & $12$ & $19$ & $\mathbf{57}$\\
\hline
\end{tabular}}

\subfigure[The number of frequency bands is $7$.]{\label{tab:compdecnf257}
\begin{tabular}{|L{0.28\columnwidth}|C{0.169\columnwidth}|C{0.169\columnwidth}|C{0.169\columnwidth}|C{0.169\columnwidth}|}
\hline
& \multicolumn{4}{c|}{Average of statistics after $100$ simulations}\\
\cline{2-5}
\theaddecnf \\
\hh
\clmnnt & $538.33$ & $\mathbf{538.74}$ & $538.31$ & $519.42$\\
\clmnns & $25.86$ & $\mathbf{25.94}$ & $25.27$ & $25.31$\\
\clmnnpd & $39.48$ & $39.33$ & $\mathbf{35.29}$ & $36.96$\\
\clmnocc & $83.76$ & $\mathbf{83.88}$ & $83.24$ & $81.53$\\
\clmncst & $\mathbf{1.68\kx^5}$ & $1.89\kx^5$ & $2.45\kx^5$ & $3.70\kx^5$\\
\clmnaoe & $3.35\kx^5$ & $\mathbf{3.28\kx^5}$ & $4.55\kx^5$ & $6.15\kx^5$\\
\hh
\disthead\\
\hh
\clmnfr & $35$ & $\mathbf{48}$ & $16$ & $1$\\
\clmnsr & $\mathbf{46}$ & $36$ & $15$ & $3$\\
\clmntr & $13$ & $10$ & $\mathbf{53}$ & $24$\\
\clmnlr & $6$ & $6$ & $16$ & $\mathbf{72}$\\
\hline
\end{tabular}}
\etb}

\section{Conclusions}\label{sec:concs}

In this work, we adapt the solution methods for the well-known machine replacement problem to the RRM problem. We propose practical performance improvement policies for the TB method. The conventional TB method does not have the capacity to adapt to the unfolding target tracking scenario. To provide some adaptation capability, we present a decision policy, DecP, and two other alternatives.

The results show that DecP based TB method yields better tracking performance, by trading the unnecessary updates of targets having accurately predictable tracks with the targets suffering from track quality degradations, say maneuvering targets. This is achieved, in effect, with the early detection of the track quality degradations via the utilization of information provided by IMM filter in the decision-making. In the numerical comparisons, it has been noted that the suggested DecP based TB method is the method with the fewest worst case tracking performance. Furthermore, the suggested policy does not only improve the average tracking performance, but can also reduce the target drops.

The suggested policy is also compared with the knowledge-based task prioritization methods based on neural networks and fuzzy logic. The neural network based scheme shows a competitive performance due to the efficient training process, while the fuzzy logic shows a rather poor performance and requires more computational time due to large number of rules. Thus, the capabilities of knowledge-based methods are limited by training process or inference rules. The suggested decision policy is rather simple and does present a good track quality improvement according to several performance metrics.

\section*{Acknowledgment}
Authors would like to thank Prof. Umut Orguner for his kind support, suggestions and insightful comments.

\def\tcon{0.28\columnwidth}
\def\tctw{0.169\columnwidth}
\def\trcon{0.04\columnwidth}
\def\trctw{0.12\columnwidth}
\def\trcth{0.12\columnwidth}
\def\trcfo{0.18\columnwidth}
\def\trcfi{0.142\columnwidth}
\def\trcsi{0.14\columnwidth}
\def\trcse{0.14\columnwidth}
\newcommand{\rcalgc}[2]{\multicolumn{1}{#1}{#2}}
\def\theadtrain{\rcalgc{|R{\trcon}|}{\centering$\#$} & \rcalgc{P{\trctw}|}{\centering Position} & \rcalgc{P{\trcth}|}{\centering Radial Velocity (m/s)} & \rcalgc{M{\trcfo}|}{Track Invalidity} & \rcalgc{P{\trcfi}|}{\centering Allowable Lateness (s)} & \rcalgc{M{\trcsi}|}{Original Priority} & \rcalgc{N{\trcse}|}{\centering Tracking Task Priority}}

\stbl{\btbdc
\caption{Training Set for Neural Network}
\label{tab:nntrain}
\begin{tabular}{|R{\trcon}|R{\trctw}|R{\trcth}|C{\trcfo}|R{\trcfi}|C{\trcsi}|L{\trcse}|}
\hline
\theadtrain \\
\hh
$1$ & $0.9939$ & $22.69$ & $0.0093$ & $0.60$ & $2$ & $0.0897$\\
$2$ & $0.7425$ & $71.45$ & $0.1832$ & $1.60$ & $4$ & $0.7205$\\
$3$ & $0.5375$ & $14.11$ & $0.0994$ & $2.20$ & $4$ & $0.6457$\\
$4$ & $0.6775$ & $50.11$ & $0.2483$ & $2.60$ & $4$ & $0.6538$\\
$5$ & $0.2919$ & $62.64$ & $0.0091$ & $2.40$ & $3$ & $0.3327$\\
$6$ & $0.5132$ & $263.77$ & $0.1026$ & $0.60$ & $1$ & $0.0862$\\
$7$ & $0.2254$ & $59.40$ & $0.1009$ & $3.00$ & $4$ & $0.6791$\\
$8$ & $0.7661$ & $203.99$ & $0.1018$ & $1.40$ & $4$ & $0.7838$\\
$9$ & $0.7286$ & $10.87$ & $0.9692$ & $3.00$ & $2$ & $0.1834$\\
$10$ & $0.0413$ & $68.90$ & $0.4899$ & $2.60$ & $1$ & $0.0653$\\
$11$ & $0.5298$ & $184.62$ & $0.9997$ & $1.20$ & $2$ & $0.5207$\\
$12$ & $0.4498$ & $103.32$ & $0.2454$ & $2.80$ & $2$ & $0.1186$\\
$13$ & $0.5654$ & $63.38$ & $0.0556$ & $1.20$ & $5$ & $0.9282$\\
$14$ & $0.5030$ & $108.20$ & $0.1226$ & $1.40$ & $2$ & $0.1599$\\
$15$ & $0.0576$ & $54.26$ & $0.0093$ & $2.20$ & $3$ & $0.4062$\\
$16$ & $0.5449$ & $220.20$ & $0.0095$ & $1.40$ & $1$ & $0.0463$\\
$17$ & $0.4213$ & $15.02$ & $0.0587$ & $2.20$ & $4$ & $0.6600$\\
$18$ & $0.4893$ & $43.44$ & $0.0919$ & $2.20$ & $5$ & $0.9018$\\
$19$ & $0.5617$ & $117.48$ & $0.0895$ & $1.20$ & $4$ & $0.7867$\\
$20$ & $0.5354$ & $10.40$ & $0.0887$ & $3.20$ & $1$ & $0.0133$\\
$21$ & $0.1073$ & $173.59$ & $0.0514$ & $2.00$ & $5$ & $0.9554$\\
$22$ & $0.4214$ & $14.72$ & $0.4778$ & $2.80$ & $1$ & $0.0337$\\
$23$ & $0.4880$ & $10.45$ & $0.1452$ & $2.00$ & $2$ & $0.1028$\\
$24$ & $0.0354$ & $212.71$ & $0.5316$ & $0.60$ & $1$ & $0.2161$\\
$25$ & $0.1571$ & $48.07$ & $0.0093$ & $2.00$ & $2$ & $0.1291$\\
$26$ & $0.0586$ & $197.90$ & $0.8060$ & $1.40$ & $1$ & $0.2234$\\
$27$ & $0.1100$ & $61.46$ & $0.0090$ & $2.40$ & $3$ & $0.3776$\\
$28$ & $0.9812$ & $51.07$ & $0.1311$ & $2.80$ & $3$ & $0.1887$\\
$29$ & $0.4913$ & $138.69$ & $1.0000$ & $2.60$ & $1$ & $0.1095$\\
$30$ & $0.4234$ & $63.91$ & $0.1501$ & $2.00$ & $4$ & $0.7409$\\
$31$ & $0.8144$ & $142.09$ & $0.1902$ & $1.60$ & $3$ & $0.4053$\\
$32$ & $0.1320$ & $11.84$ & $0.2947$ & $3.20$ & $2$ & $0.1159$\\
$33$ & $0.6585$ & $154.45$ & $0.2427$ & $1.60$ & $1$ & $0.0449$\\
$34$ & $0.0717$ & $54.23$ & $0.1179$ & $3.00$ & $3$ & $0.3649$\\
$35$ & $0.8461$ & $137.78$ & $0.9992$ & $0.60$ & $3$ & $0.7902$\\
\hline
\end{tabular}
\begin{tabular}{|R{\trcon}|R{\trctw}|R{\trcth}|C{\trcfo}|R{\trcfi}|C{\trcsi}|L{\trcse}|}
\hline
\theadtrain \\
\hh
$36$ & $0.6705$ & $159.16$ & $0.0369$ & $0.80$ & $5$ & $0.9471$\\
$37$ & $0.9792$ & $150.03$ & $0.0669$ & $1.60$ & $5$ & $0.9026$\\
$38$ & $0.8751$ & $61.95$ & $0.0686$ & $2.20$ & $5$ & $0.8597$\\
$39$ & $0.6780$ & $63.46$ & $0.0499$ & $2.60$ & $4$ & $0.5866$\\
$40$ & $0.6825$ & $51.20$ & $0.0095$ & $1.80$ & $1$ & $0.0203$\\
$41$ & $0.6074$ & $108.30$ & $0.2895$ & $2.00$ & $5$ & $0.9372$\\
$42$ & $0.8189$ & $33.63$ & $0.0427$ & $2.40$ & $5$ & $0.8401$\\
$43$ & $0.9859$ & $83.21$ & $0.6677$ & $3.00$ & $3$ & $0.3632$\\
$44$ & $0.8068$ & $10.13$ & $0.2872$ & $2.80$ & $3$ & $0.2428$\\
$45$ & $0.3788$ & $41.13$ & $0.7067$ & $2.20$ & $2$ & $0.2482$\\
$46$ & $0.8867$ & $62.63$ & $0.0093$ & $2.40$ & $2$ & $0.0555$\\
$47$ & $0.9127$ & $30.27$ & $0.0930$ & $2.60$ & $4$ & $0.5146$\\
$48$ & $0.6896$ & $82.32$ & $0.0587$ & $2.60$ & $5$ & $0.8694$\\
$49$ & $0.7627$ & $61.15$ & $0.5034$ & $2.80$ & $3$ & $0.3608$\\
$50$ & $0.2818$ & $66.24$ & $0.0094$ & $2.60$ & $1$ & $0.0235$\\
$51$ & $0.7048$ & $60.13$ & $0.3315$ & $3.00$ & $3$ & $0.2932$\\
$52$ & $0.5438$ & $34.47$ & $0.1127$ & $3.20$ & $4$ & $0.5637$\\
$53$ & $0.0727$ & $74.72$ & $0.0879$ & $1.40$ & $2$ & $0.2061$\\
$54$ & $0.9995$ & $35.23$ & $0.1029$ & $3.20$ & $3$ & $0.1488$\\
$55$ & $0.1679$ & $102.10$ & $0.0669$ & $1.40$ & $5$ & $0.9549$\\
$56$ & $0.3996$ & $229.30$ & $1.0000$ & $0.60$ & $3$ & $0.8915$\\
$57$ & $0.4460$ & $190.43$ & $0.1017$ & $1.60$ & $3$ & $0.5068$\\
$58$ & $0.1662$ & $116.84$ & $0.0529$ & $0.60$ & $4$ & $0.8734$\\
$59$ & $0.7138$ & $46.85$ & $0.3064$ & $2.60$ & $5$ & $0.8972$\\
$60$ & $0.4875$ & $20.41$ & $0.4455$ & $2.60$ & $1$ & $0.0330$\\
$61$ & $0.7214$ & $62.43$ & $0.0644$ & $2.20$ & $5$ & $0.8784$\\
$62$ & $0.3649$ & $76.44$ & $0.5753$ & $2.60$ & $1$ & $0.0545$\\
$63$ & $0.9461$ & $42.89$ & $0.0906$ & $2.40$ & $5$ & $0.8357$\\
$64$ & $0.2150$ & $103.39$ & $0.1191$ & $1.80$ & $1$ & $0.0466$\\
$65$ & $0.6542$ & $171.75$ & $0.9799$ & $1.60$ & $4$ & $0.9355$\\
$66$ & $0.2706$ & $36.96$ & $0.0381$ & $2.00$ & $5$ & $0.9194$\\
$67$ & $0.2538$ & $105.43$ & $0.0093$ & $2.40$ & $2$ & $0.1188$\\
$68$ & $0.3348$ & $119.44$ & $0.1797$ & $2.00$ & $4$ & $0.7975$\\
$69$ & $0.8881$ & $313.40$ & $0.0730$ & $0.60$ & $2$ & $0.2348$\\
$70$ & $0.1553$ & $99.41$ & $0.9987$ & $1.60$ & $2$ & $0.5140$\\
\hline
\end{tabular}
\etbdc}

\appendix[Descriptions of the Task Prioritization Methods Evaluated for Comparisons\label{app}]
Task prioritization methods based on neural network \cite{komorniczak,komorniczak2} and fuzzy logic \cite{miranda} could not be implemented directly as described in references, due to distinctions between problem models. We define common input variables to adapt these methods to our radar system and to fairly compare them. Input variables are as follows:
\begin{itemize}
	\item \textit{Position} variable denotes the normalized position of a target relative to the innermost point of a priority ring within the same azimuth angle, e.g. the position of a target at the range of $50$ km is $\operatorname{mod}(50,40)/40=0.25$. It varies between $0$ and $1$.
	\item \textit{Radial velocity} variable denotes the radial velocity of a target. For evaluations, it varies between $0$ and $350$ m/s. Hence, the radial velocity higher than $350$ m/s is truncated to $350$ m/s.
	\item \textit{Tracking invalidity} variable denotes the severity of average tracking error. We utilize the hyperbolic tangent sigmoid transfer function $\operatorname{tansig}(\cdot)$, which is mathematically equivalent to the hyperbolic tangent function $\operatorname{tanh}(\cdot)$, such that the tracking invalidity of a target is
	\begin{equation}
	\operatorname{TI}(x) = \dfrac{2}{1+e^{-2x/{10}^6}}-1,
	\end{equation}
	where $x$ is the average tracking error of target. Since $x$ is a non-negative number, $\operatorname{TI}(x)$ varies between $0$ and $1$. The smaller error makes the tracking invalidity smaller.
	\item \textit{Allowable lateness} is a tolerable time difference between actual update time at which the task can be scheduled and due time by which it must be scheduled to successfully accomplish for late update. It varies between $0.6$ and $3.2$ s. The allowable lateness longer than $3.2$ s is truncated to $3.2$ s.
	\item \textit{Original priority} is the priority assigned by the operator. Within the range of $200$ km, it is decreased from $5$ to $1$ by $1$ through each ring having $40$ km thickness.
\end{itemize}
\textit{Tracking task priority} is the output variable changing between $0$ and $1$.

\begin{figure*}[!t]
	\centering
	\subfigure[]{
		\includegraphics{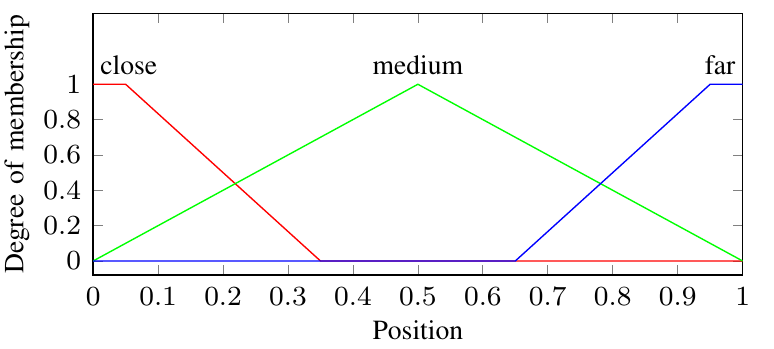}
		\label{fig:fltpmfi1}
	}
	\subfigure[]{
		\includegraphics{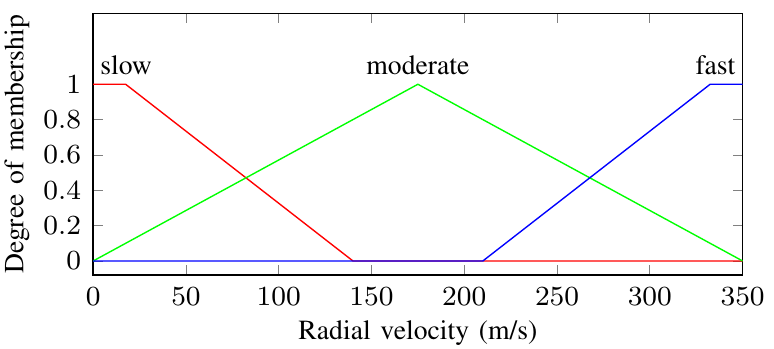}
		\label{fig:fltpmfi2}
	}
	\subfigure[]{
		\includegraphics{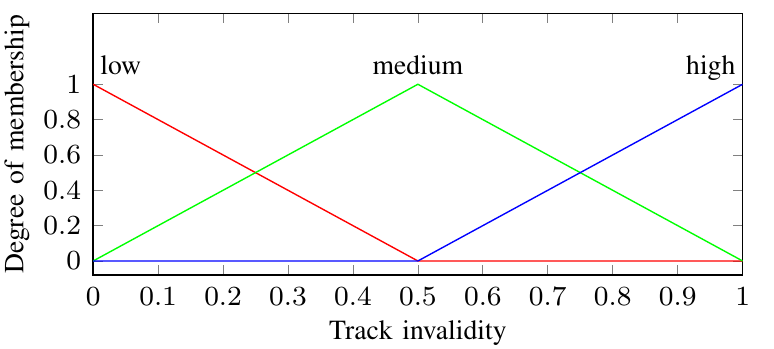}
		\label{fig:fltpmfi3}
	}
	\subfigure[]{
		\includegraphics{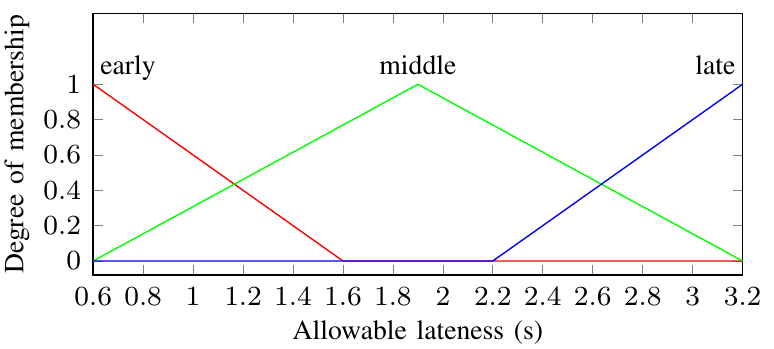}
		\label{fig:fltpmfi4}
	}
	\subfigure[]{
		\includegraphics{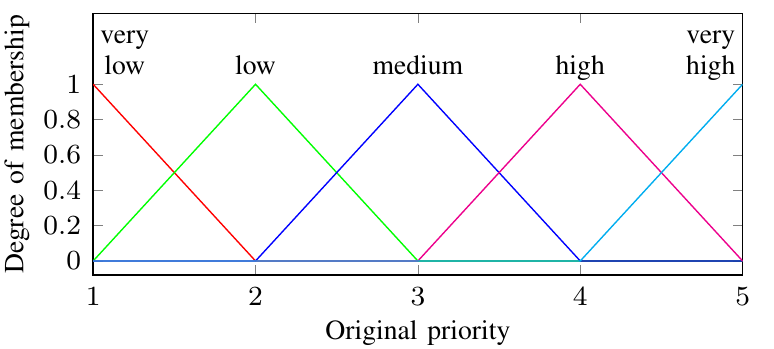}
		\label{fig:fltpmfi5}
	}
	\subfigure[]{
		\includegraphics{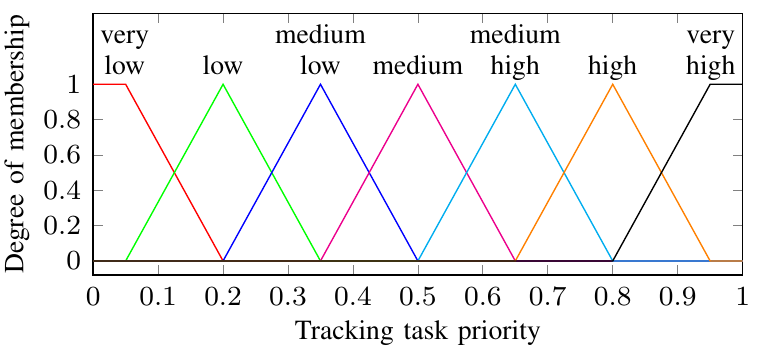}
		\label{fig:fltpmfo}
	}
	\caption{Membership functions of fuzzy input variables, \subref{fig:fltpmfi1} position, \subref{fig:fltpmfi2} radial velocity, \subref{fig:fltpmfi3} track invalidity, \subref{fig:fltpmfi4} allowable lateness and \subref{fig:fltpmfi5} original priority; fuzzy output variable, \subref{fig:fltpmfo} tracking task priority.}
	\label{fig:fltpmf}
\end{figure*}

\begin{figure}[!t]
	\centering
	\includegraphics{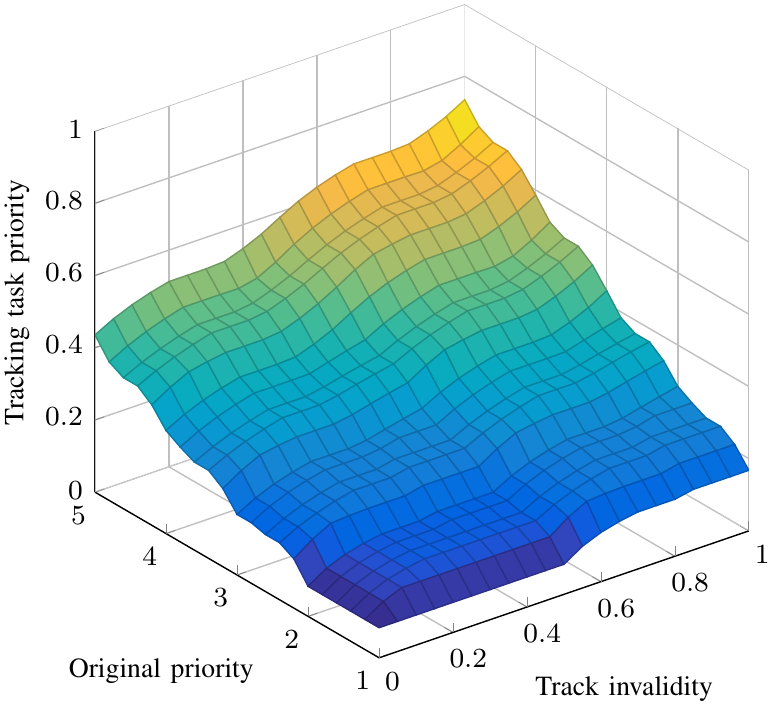}
	\caption{Surface representation of the fuzzy logic task prioritization system for fixed input variables, position, radial velocity and allowable lateness.\label{fig:fltpmfi35}}
\end{figure}

\subsection{Neural Network}
We have trained the network with training set given in Table~\ref{tab:nntrain}. These data are taken from varying simulations with underloaded and overloaded conditions. There is one hidden layer with $5$ neurons. Transfer function of the hidden layer is $\operatorname{tansig}$ and the output layer is $\operatorname{logsig}$.
\subsection{Fuzzy Logic}
The membership functions of fuzzy values corresponding to fuzzy variables are shown in Fig.~\ref{fig:fltpmf}. We have defined $333$ fuzzy rules based on these fuzzy values. In Fig.~\ref{fig:fltpmfi35}, the surface representation of tracking task priority is shown when the input variables, position, radial velocity and allowable lateness are fixed as $0.5$, $175$ m/s and $1.2$ s, respectively.


%

\begin{IEEEbiography}[{\includegraphics[width=1in,height=1.25in,clip,keepaspectratio]{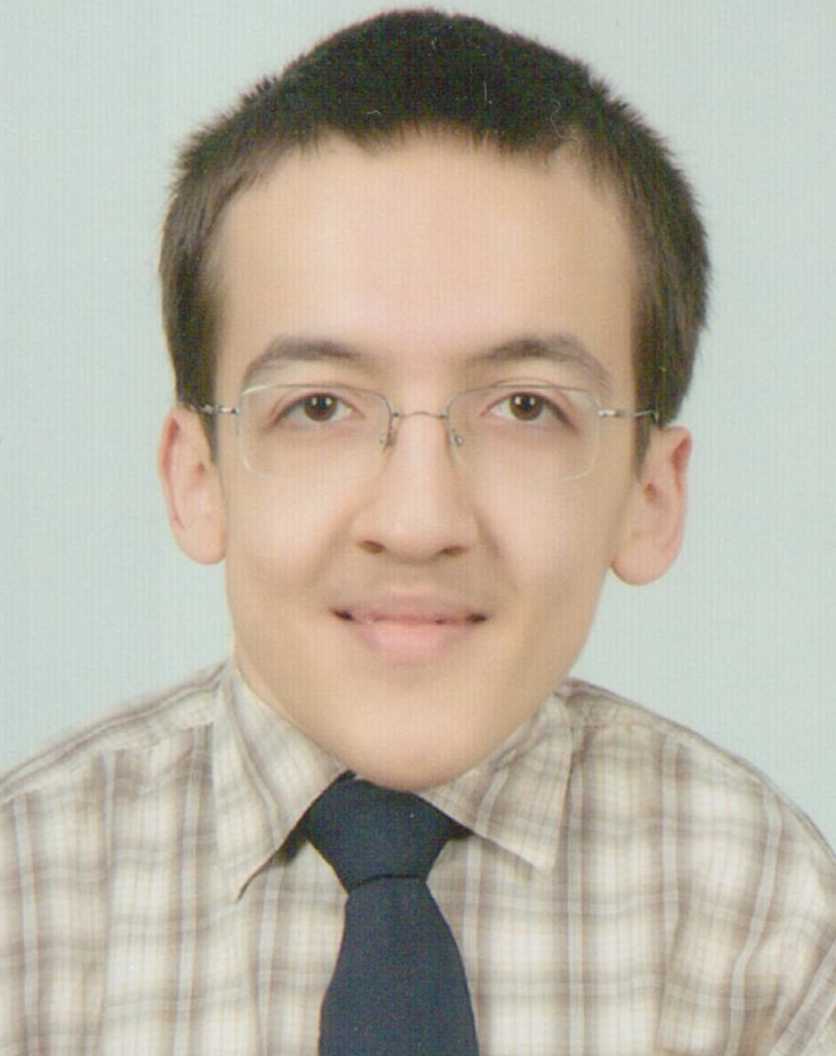}}]{\"Omer \c{C}ay{\i}r}
is a research assistant  pursuing Ph.D. degree at the Department of Electrical \& Electronics Engineering of Middle East Technical University, Ankara, Turkey. All in electrical and electronics engineering, he received the B.S. degree (with rank 1) from Hacettepe University, Ankara, Turkey in 2011 and M.S. degree from Middle East Technical University, Ankara, Turkey in 2014.
	
His research interests include statistical signal processing and its applications in waveform optimization, stochastic control, radar resource management and software defined radio.
\end{IEEEbiography}

\begin{IEEEbiography}[{\includegraphics[width=1in,height=1.25in,clip,keepaspectratio]{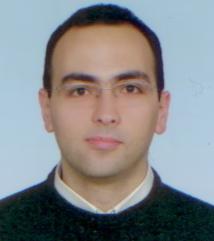}}]{\c{C}a\u{g}atay Candan}
is a professor at the Department of Electrical \& Electronics Engineering of Middle East Technical University, Ankara, Turkey. He received his B.S., M.S., and Ph.D. degrees, all in electrical engineering, from Middle East Technical University, Ankara, Turkey (1996), Bilkent University, Ankara, Turkey (1998) and Georgia Institute of Technology, Atlanta, USA (2004), respectively.
	
His research interests include statistical signal processing and its applications in array signal processing, radar signal processing and communications.
\end{IEEEbiography}





\end{document}